\documentclass[12pt]{article}

\usepackage[dvips]{graphics}
\usepackage{epsfig}
\usepackage{wrapfig}
\usepackage{color}
\usepackage{xspace}
\usepackage{epsfig} 

\usepackage{enumitem}

\usepackage{amsmath,amssymb,amsfonts,bm,amsthm}
\usepackage{graphicx,rotating,lscape}
\usepackage{hyperref}
\usepackage{natbib} 
\usepackage{booktabs} 
\usepackage[all]{xy}
\usepackage{color}
\usepackage{latexsym}
\usepackage{authblk}
\usepackage{soul}
\usepackage{lineno}

\newtheorem{lemma}{Lemma}       
\newtheorem{theorem}{Theorem}      
\newtheorem{proposition}{Proposition}
\newtheorem{definition}{Definition}

\newtheorem{example}{Example}

\def\real{I\!\!R}
\newcommand{\bX}{{\mathbf X}}

\newcommand{\pr}{P}

\newcommand{\btheta}{{\mbox{\boldmath $\theta$}}}

\definecolor{cambridgeblue}{rgb}{0.64, 0.76, 0.68}
\definecolor{camouflagegreen}{rgb}{0.47, 0.53, 0.42}

\definecolor{capri}{rgb}{0.0, 0.75, 1.0}
\definecolor{caribbeangreen}{rgb}{0.0, 0.8, 0.6}

\newcommand{\Minge}[1]{{\color{black}{#1}}}

\title{Nonparametric fusion learning: synthesize inferences from diverse sources using depth confidence distribution}
\author{\large \emph{}}

 \author[1]{Dungang Liu}
\author[2]{Regina Y. Liu}
\author[2]{Minge Xie}
\affil[1]{Department of Operations, Business Analytics and Information Systems, University of Cincinnati, Cincinnati, Ohio 45221, USA}
\affil[2]{Department of Statistics and Biostatistics, Rutgers University, Piscataway, New Jersey 08854, USA}
\date{}

\begin{document}
\renewcommand{\arraystretch}{1.1}
\baselineskip = 7mm
\parskip = 0mm 

\maketitle
\vspace{-0.5in}
\begin{abstract}
Fusion learning refers to synthesizing inferences from multiple sources or studies to provide more effective inference and prediction than from any individual source or study alone. Most existing methods for synthesizing inferences rely on parametric model assumptions, such as normality, which often do not hold in practice. In this paper, we propose a general nonparametric fusion learning framework for synthesizing inferences of the target parameter from multiple sources. 
The main tool underlying the proposed framework is the notion of {\it depth confidence distribution ({\it depth-CD})}, which is also developed in this paper. Broadly speaking, a {\it depth-CD} is a data-driven nonparametric  summary distribution of  inferential information for the target parameter. We  show that a {\it depth-CD} is a useful inferential tool and, moreover, is an omnibus form of confidence regions (or $p$-values),  whose contours of level sets shrink toward the true parameter value. The proposed fusion learning approach combines {\it depth-CD}s from the individual studies, with each {\it depth-CD} constructed by nonparametric bootstrap and data depth. This approach is shown to be {\it efficient}, {\it general} and {\it robust}. Specifically, it achieves high-order accuracy and Bahadur efficiency under suitably chosen combining elements. It allows the model or inference structure to be different among individual studies. And it readily adapts to heterogeneous studies with a broad range of complex and irregular settings. This property enables it to utilize indirect evidence from incomplete studies to gain efficiency in the overall inference.
In addition to developing the theoretical support for the proposed approach, we also 
apply the approach to making combined inference for the common mean vector and correlation coefficient from several studies. The numerical results from simulated studies show the approach to be less biased and more efficient than the traditional approaches in non-normal settings. The advantages of the proposed approach are also demonstrated in a Federal Aviation Administration (FAA) study of aircraft landing performance.
\end{abstract}

Key words: common parameter, evidence synthesis, fusion learning, heterogeneous studies, meta-analysis, multiparameter inferences, $p$-value function.

\footnotetext[1]{Dungang Liu is Associate Professor (Email: dungang.liu@uc.edu), Department of Operations, Business Analytics and Information Systems, University of Cincinnati Lindner College of Business, Cincinnati, Ohio 45221. Regina Y. Liu and Minge Xie are Distinguished Professors,(Emails:rliu@stat.rutgers.edu and mxie@stat.rutgers.edu), Department of Statistics, Rutgers University, New Brunswick, NJ 08903. Their research is supported in part by the NSF grants DMS1737857, DMS1812048, DMS2015373 and DMS2027855.}

\newpage
\section{Introduction}
Powerful data acquisition technologies have greatly enabled the simultaneous collection of data from different sources in many domains. It is often impossible or inappropriate to simply aggregate all the data to draw inference, due to concerns over storage, privacy or cost constraints, or the desire to enhance inference by incorporating external or publicly available data sources, etc.  Instead, one would need to combine the inference results from individual sources to draw an overall inference. Fusion learning refers to synthesizing inferences from multiple sources or studies to provide a more effective inference than that from any individual source or study alone. 

\begin{figure}
    \includegraphics[width=16cm, height = 7cm]{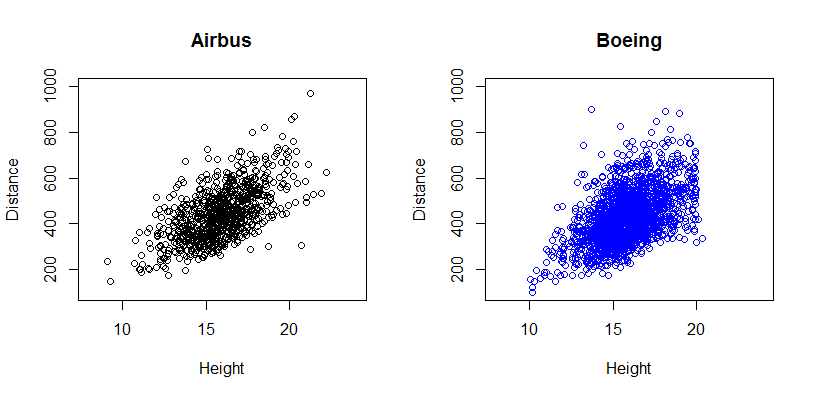}
    \caption{Scatter plots of Distance versus Height for the two aircraft makes, {\it Airbus} and {\it Boeing}.}
    \label{fig:FAA0}
\end{figure}

\medskip
\noindent{$\bullet$ \bf A motivating example}\\
We begin with an example to illustrate the need of efficient fusion learning. This example arose from a research project sponsored by the Federal Aviation Administration (FAA). The FAA, as the regulatory agency for air transportation safety, establishes guidelines for all air operations. For example, to ensure safe aircraft landings,  FAA analysis has set  guidelines recommending that the height of the aircraft at the crossing of runway threshold  be around 15.85m  and touchdown distance be around 432m from runway threshold. To help assess whether aircraft landings generally follow these guidelines, we can simply test the hypothesis $H_0: \bm\mu =(15.85, 432)^\prime$, where $\bm\mu$ is the mean vector for the height and distance. A sample of 2796 landing records (820 are from {\it Airbus} and 1976 from {\it Boeing}) yields a sample mean of $(15.86, 432.95)^\prime$, and a $p$-value of 0.942 from Hotelling's $T^2$ test. The finding would lead to the conclusion that aircraft landings generally comply with the FAA guidelines. Surprisingly, this conclusion appears to contradict the conclusion that we would draw from the two separate individual tests from {\it Airbus} and {\it Boeing}, with respective $p$-values 0.006 and 0.167.
A closer examination of the scatter plots in Figure~\ref{fig:FAA0}, of the two individual studies for {\it Airbus} and {\it Boeing}, indicates that the two samples do not appear to follow the same distribution and neither follows an elliptical distribution, and that the {\it Boeing} sample seems to be truncated on the right. This  casts doubt on the aforementioned conclusion of landing operations meeting the FAA guidelines, 
and suggests the need of a nonparametric test for the hypothesis that both landing operations from {\it Airbus} and {\it Boeing} meet the FAA guidelines, i.e.,  $H: \mu_{Airbus} =\mu_{Boeing}=(15.85, 432)$. More importantly, this example shows that blindly aggregating data from different data sources may not necessarily yield correct overall inferences. This example is discussed further in Section~\ref{sec:FAA-example}.

\medskip
\noindent{$\bullet$ \bf Outline of the proposed approach and highlights of results}\\
In this paper, we develop an efficient nonparametric approach for fusion learning to make inference for the common parameters shared by different studies. 
This approach consists of two key parts: a) We develop a general nonparametric inference procedure to ascertain a valid inference for each individual study by applying the notion of  {\it depth confidence distribution} ({\it depth-CD}) and its associated {\it depth confidence curve} ({\it depth-CV}). Specifically in this paper, we construct a {\it depth-CD} using {\it data depth} and {\it bootstrap} and show the {\it depth-CD} as a 
comprehensive summary distribution of all the inferential information for the target parameter;  b) We derive an overall combined inference by suitably combining the {\it depth-CV}s from the individual studies.
Our proposed approach for individual-study inference and that for the combined  inference are completely nonparametric and data driven, and broadly applicable without any model assumptions. For instance, this can substantially broaden the scope of the existing meta-analysis and evidence synthesis, where common practice routinely requires parametric models, often the normality assumption, see, e.g., \citep{normand1999meta,sutton2008recent}. 

The proposed fusion learning framework is established based on {\it depth confidence distribution} ({\it depth-CD}), which is a new powerful inference tool developed in this paper by using three distinct concepts: confidence distribution \citep[e.g.,][]{xie2013confidence,schweder2016confidence}, data depth \citep[e.g.,][]{liu1990notion,  liu1999multivariate, zuo2000general}, and bootstrap \citep{Efron79}. Simply put,
a {\it depth-CD} is a sample-dependent distribution function defined on the parameter space which  summarizes the information from the data that is relevant for the inference of parameters. Based on the evidence in the given data, a {\it depth-CD} can also be viewed as a reference function that reflects the plausibility or ``confidence'' associated with each possible parameter value on the parameter space. We investigate general properties of {\it depth-CD}, in particular the following three, in Sections 3.2-3.4,\\
\vspace{-8mm}
\begin{itemize}[itemsep=-4pt]
 \item[($\mathcal{P}$-1)] a {\it depth-CD} is an omnibus form of confidence regions at all confidence levels;
     \item[($\mathcal{P}$-2)] a {\it depth-CD} is an omnibus form of $p$-values  for testing values on the entire parameter space;
     \item[($\mathcal{P}$-3)] the contours of the level sets of a {\it depth-CD} shrink toward the true value of the parameter.
\end{itemize}
\noindent
These properties show that a {\it depth-CD} is useful in yielding all inference outcomes commonly sought in practice, and also that  it is  a versatile tool for nonparametric fusion learning.

Under the proposed general {\it depth-CD}  fusion learning framework, 
 we develop an efficient nonparametric fusion learning approach by fusing the {\it depth-CD}s from individual studies where the {\it depth-CD} of each study is constructed from data depth and nonparametric bootstrap as described in Section~\ref{sec:CD-bootstrap}. The fused output, similar to the individual input, remains a distribution function on the parameter space, which now depicts the level of ``confidence'' in assuming each possible parameter value in view of the totality of all available evidence gathered from all studies. This combined {\it depth-CD}, following $\mathcal{P}$-1,2,3 above, can readily provide an overall inference as confidence regions, $p$-values, or consistent point estimators. 

The proposed fusion approach is shown to be {\bf efficient}, {\bf general} and {\bf robust}. More precisely, it is {\bf efficient}, as it achieves high-order accuracy and Bahadur efficiency under suitably chosen combining elements, as shown in Section 5.1. It is {\bf general}, as a) it covers multiparameter settings, b) it is nonparametric, and c) it permits flexible choices of transformations of input functions, weighting schemes and methods for deriving each individual {\it depth-CD}, across all studies. Such choices are often needed to account for the different circumstances or degrees of trustworthiness surrounding each individual study. It is {\bf robust}, as it adapts efficiently to the fusion of heterogeneous studies, covering a wide range of complex and irregular studies, as investigated in Section 5.2. In fact, our fusion approach covers the particularly challenging setting where the target parameter may not be even estimable in some subset of studies, such as in the case of incomplete studies.  
Although the target parameter vector may not be estimable in incomplete studies, those studies often contain information from their data that can contribute to the overall inference of the target parameter, as the information among different component parameters is often correlated, see, e.g., \cite{liu2015multivariate}. This data information from incomplete studies is often regarded as {\it indirect evidence}. Therefore, our fusion approach can incorporate both direct and indirect evidence, all in a nonparametric manner. This is a desirable property since it gains efficiency in the overall inference, as shown in Section 6.1.
 
We present an extensive comparison study of our fusion method with several existing methods in the setting of making inference for a common mean vector, in three data scenarios.  
The results can be summed up as three advantages of our method, namely, in the absence of the normality assumption: 1) it preserves inference accuracy in hypothesis testing/confidence regions; 2) its point estimator has less bias and is more efficient; and 3) it achieves a gain of efficiency in the presence of heterogeneous studies. We also present a numerical study of our method in meta-analysis of correlation coefficients in Section 6.2. 
There we observe that traditional methods may yield misleading conclusions, while ours remains valid in both normal and non-normal cases.

The remaining paper is organized as follows. Section~2  gives a general setup for fusion learning.
Section 3 covers a brief review of confidence distribution (CD) and data depth, and then the development of  {\it depth-CD} and depth confidence curve ({\it depth-CV}) for multiparameter inference. Section 4 provides a concrete procedure for constructing a {\it depth-CV} by using bootstrap and data depth. Section 5 develops  nonparametric fusion learning by combining the {\it depth CD}s derived from individual studies. Section 6 covers all simulation studies. Section 7 applies our fusion learning approach to conduct the FAA study of aircraft landing performance. Section~8 contains more comments and discussions.

\section{A general problem setting for fusion learning}
We consider the problem of fusion learning in a general setting. Suppose that $K$ independent studies are available  for analysis to address the same scientific or business question. Let
\begin{equation}
  \bm X_{k,1},\bm X_{k,2},\ldots,\bm X_{k,n_k}, \ i.i.d.\ \sim \mathcal{F}_k,
  \label{model}
\end{equation}
be the sample from the $k$-th study, where $\mathcal{F}_k$ is an unknown $p_k$-dimensional multivariate distribution. Assume that  the parameter of interest $\bm\theta_k$ is a finite-dimensional functional of $\mathcal{F}_k$, which can be scalar or vector-valued. Assume that
\begin{equation}
  \label{eq:assumption-homo}
  \bm\theta \equiv \bm\theta_1=\bm\theta_2=\cdots=\bm\theta_K.
\end{equation}
The goal is to make an efficient inference for $\bm\theta$ by fusing the information from all $K$ studies, without assuming specific parametric forms of the distributions $\mathcal{F}_k(\bm\theta_k)$. This setting covers:

\begin{example}[common mean inference]
  Let $\bm\theta_k=\int {\bf x}~\mathrm{d}\mathcal{F}_k({\bf x})$ be the mean of the distribution $\mathcal{F}_k$, $\bm\theta \equiv \bm\theta_1=\cdots=\bm\theta_K$ is the ($p$-dimensional) common mean of the $K$ unknown distributions. We are interested in constructing a confidence region for $\bm\theta$ or testing the hypothesis $\Sigma_0: \bm\theta=\bm\mu~\text{versus}~\Sigma_1:\bm\theta \neq \bm\mu$ for a particular value $\bm\mu$.
\end{example}
\begin{example}[correlation inference]
  Consider the correlation coefficient of any two components of the $p$-dimensional distribution $\mathcal{F}_k$. Let $\bm\theta_k$ include all such pairwise correlation coefficients. Then $\bm\theta \equiv \bm\theta_1=\cdots=\bm\theta_K$ is a parameter vector of dimension $p(p-1)/2$. We are  interested in testing the hypothesis $\Sigma_0: \bm\theta=\bm0~\text{versus}~\Sigma_1:\bm\theta \neq \bm0$.
\end{example}

We will use these two as illustrative examples throughout the development and simulations of the proposed fusion approach. These examples in various scenarios can help showcase the merits of our approach, described briefly as {\bf efficient}, {\bf general} and {\bf robust} in the {\bf Introduction}.
To elaborate further, 
our fusion framework is general because it requires no specific parametric forms of the underlying distributions $\mathcal{F}_k$.  It is also robust because it permits a broad range of heterogeneity among studies: i) the individual studies do not have be homogeneous in terms of their designs, reporting formats, models, and inference methods.; ii) the studies can have different types of data (e.g., continuous, binary or ordinal responses);  iii) the studies can be analyzed using different models, such as linear regression models for continuous outcomes in some studies and logistic regression models for binary outcomes in others, and iv) the individual studies can  even use different inference methods, for instance, estimating the population location by the sample mean, the trimmed mean or the median as dictated by the specific situation of each study; and v) our fusion framework does not require that the parameter $\bm\theta_k$ be estimable in all studies. 
 
To formulate the last point precisely, our CD fusion approach applies even if the parameter of interest $\bm\theta_k$ is not estimable in some studies, as long as there exists a continuous mapping from the parameter space $\bm \Theta$  (of $\bm\theta$) to a lower-dimensional space $\bm\Theta_k$ such that
\begin{equation}
  \label{eq:assumption-heter}
  \tilde{\bm\theta}_k=\bm f_k(\bm\theta_k)
\end{equation}
is estimable.  Similar formulation of partially estimable parameters also arose in the applications  in \citep{sutton2008recent,liu2015multivariate}  Obviously, when all $\bm f_k$'s are identity mappings, this setting reduces to the case where all $\bm \theta_k$'s 
are estimable. Our fusion approach is thus adaptable to such indirect evidence. Two numerical examples in Sections 6.1 and 7  illustrate how our approach gains efficiency in the final combined inference from  incorporating indirect evidence.

\section{depth-CD and depth-CV for multiparameter inference}
\label{sec:depth-CD-inference}

\subsection{Reviews: Confidence distribution (CD) and data depth}
\subsubsection{Confidence distribution (CD) and confidence curve (CV) for scalar parameter}\label{sec:CD}
The idea of the confidence distribution (CD) is borne out of the wish to use a sample-dependant distribution function, rather than a point estimate or an (confidence) interval estimate, to estimate an unknown parameter. 
For a scalar parameter $\theta \in \Theta$, a function $H_n(\cdot)\equiv H_n(\bm X_n,\cdot)$ is said to be a CD function for $\theta$ if it meets these two requirements: (i) given a sample $\bm X_n$, it is a distribution function on $\Theta$; and (ii) at the true parameter value $\theta=\theta^o$, $H_n(\theta^o)\equiv H_n(\bm X_n,\theta^o)$, as a function of the sample $\bm X_n$, follows the uniform distribution on (0,1) \Minge{\citep{schweder2002confidence, singh2005combining}}. Essentially, (i) says that a CD function is  a {\it 'distribution estimate'} dependent on the observed sample, and (ii) ensures that a CD function carries frequentist properties in terms of repeated sampling. For instance, under (ii), $(-\infty, H_n^{-1}(1-\alpha))$ is a $(1-\alpha)$ confidence interval, and also $H_n(\theta^o)$ can be used as a $p$-value for testing the hypotheses $\Omega_0: \theta \leq \theta^o$ versus $\Omega_1: \theta > \theta^o$. More precisely, given a data set and an inferential procedure, a CD function represents a set of confidence intervals for all possible confidence levels. It describes a {\it distribution of confidence} associated with  each $\theta$ value in $\Theta$.

Note that, conditional on the observed sample $\bX_n$, a CD function $H_n(\theta)\equiv H_n(\bm X_n,\theta)$ is a distribution function on the parameter space $\Theta$. Let $\theta^*$ be a random variable following  the distribution $H_n(\cdot)$. We refer to $\theta^*$ as a {\it CD-random variable}. Conditional on the given data, we can used simulated samples $\theta^*$'s from $H_n(\cdot)$ to carry out inference, as discussed in Section~\ref{sec:CD-bootstrap}. 

\begin{figure}
    \includegraphics[width=16cm, height = 7cm]{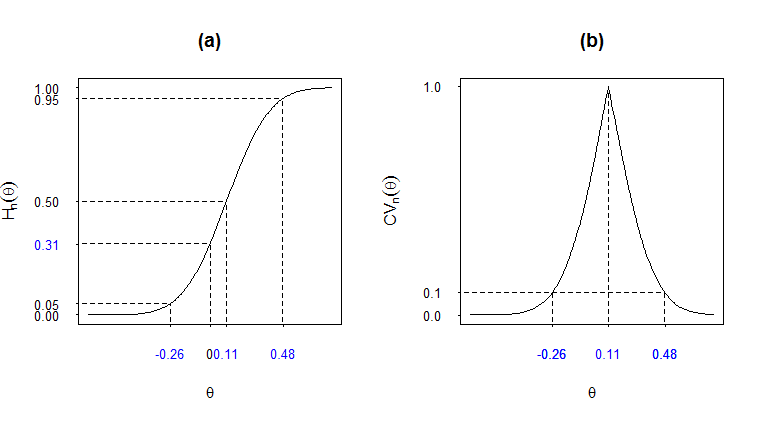}
    \caption{The curves represent a confidence distribution function (a) and the corresponding confidence curve (b) for the mean parameter $\theta$ in the normal distribution $N(\theta,1)$. They are obtained based on a  sample $\bm{x}=\{x_i, i=1,\ldots,20\}$ from $N(0,1)$. Illustrated is how to draw commonly used inferential outcomes such as a point estimate of 0.11, a 90\% confidence interval of (-0.26, 0.48), and a $p$-value of 0.31 for testing the hypothesis $\Omega_0:\theta\le 0$ versus $\Omega_1:\theta>0$.}
    \label{fig:CD-CV-Univariate}
\end{figure}

To illustrate the CD inference approach, we consider the simple example with a sample $\bm{x}=\{x_i, i=1,\ldots,n\}$ from $N(\theta,1)$, where the mean $\theta$ is the parameter of interest. A natural CD for $\theta$ is $\mathcal{N}(\bar{x}_n,1/n)$ or equivalently its cumulative distribution function $H_n(\theta)=\Phi(\sqrt{n}(\theta-\bar{x}_n))$. Given a sample $\bm{x}$, the function  $H_n(\theta)$ is a distribution function on the parameter space $\Theta$, and it carries all commonly used inference outcomes. For instance, $(H_n^{-1}(\alpha/2),H_n^{-1}(1-\alpha/2))=(\bar{x}_n+\Phi^{-1}(\alpha/2)/\sqrt{n},\bar{x}_n+\Phi^{-1}(1-\alpha/2)/\sqrt{n})$ is a $(1-\alpha)$ confidence interval for $\theta$, for any $0<\alpha\le1$; the mean/median of $(H_n^{-1}(.5)$ (= $\bar{x}_n$) is a point estimate for $\theta$; and the tail mass $H_n(b)=\Phi(\sqrt{n}(b-\bar{x}_n))$) is a $p$-value for testing the one-sided hypothesis $K_0:\theta\le b$ versus $K_1:\theta>b$. The curve in Figure~\ref{fig:CD-CV-Univariate}(a) is a CD function given a random sample of size $n=20$. The dashed lines there help illustrate all types of inference outcomes from a CD function, including a point estimate of 0.11, a 90\% confidence interval of (-0.26, 0.48), and a $p$-value of 0.31 for testing $\Omega_0:\theta\le 0$ versus $\Omega_1:\theta>0$.

For the ease of visualization of confidence intervals of different levels, the distributional form of a CD  $H_n(\cdot)\equiv H_n(\bm X_n,\cdot)$ seen in in Figure~\ref{fig:CD-CV-Univariate}(a) can be expressed alternatively as a confidence curve (CV) seen in Figure~\ref{fig:CD-CV-Univariate}(b) which is defined as
\begin{equation}
  \label{eq:CV}
CV_n(\theta)=1-2 | H_n(\theta)-0.5 | = 2 \min\{H_n(\theta), 1-H_n(\theta)\},
\end{equation}
(see \citealp{xie2013confidence} pp.29-31; \citealp{schweder2016confidence} pp.10-14). 
While the CD function $H_n(\theta)$ represents the upper limits of one-sided confidence intervals, the confidence curve $CV_n(\theta)$ gives an omnibus form of the limits of two-sided confidence intervals. 
In Figure~\ref{fig:CD-CV-Univariate}(b), the two limits of a 90\% confidence interval identified by the two points on the confidence curve at the height $\alpha=0.1$ are exactly the same as those obtained from Figure~\ref{fig:CD-CV-Univariate}(a).
Furthermore, following the duality between confidence intervals and hypothesis testing, $CV_n(\theta^0)$ can serve as a $p$-value function for the two-sided hypothesis testing, $\Omega_0: \theta = \theta^o$ versus $\Omega_1: \theta \neq \theta^o$, for any $\theta^o \in \Theta$. Also, the confidence curve peaks at the median of the CD function, i.e., $CV_n^{-1}(1)=0.11$ as shown in Figure~\ref{fig:CD-CV-Univariate}(b), which yields a median-unbiased estimate for $\theta$.  

Without linking to CD, the concept of CV has actually been explored in \citep{birnbaum1961confidence,blaker2000confidence,blaker2000paradoxes} for a scalar parameter $\theta$. In fact, 
\cite{blaker2000paradoxes} interprets a CV as a summary of ``how each parameter value is ranked in view of the data'' from the peak decreasing gradually along the tails. This ranking interpretation of the CV in fact suggests a natural extension of the CD to the multiparameter setting by incorporating the notion of data depth, which has been developed to establish a center-outward ordering of multivariate observations. We will develop this extension after the brief review of data depth and its properties.

\subsubsection{Data depth and center-outward ordering of multivariate data}

{\it Data depth} is a way to measure how deep or central a given point is with respect to a multivariate sample cloud, say $\{\bm\xi_1,...,\bm\xi_m\} \sim F \in \real^p$, or to its underlying distribution $F$. It  naturally yields a measure of ``outlyingness'' and thus also a center-outward ordering of these multivariate points.
Common depth functions include, Mahalanobis depth (MD) \citep{mahalanobis1936generalized},
half-space depth (HD) \citep{hodges1955bivariate,tukey1975mathematics},
 simplicial depth (SD) \citep{liu1990notion},
among others.

Using simplicial depth as an example, the depth at point ${\bf z} \in \real^p$ with respect to $F$ is $D_F({\bf z})=P_F\{{\bf z}\in s[\bm\xi_1,\ldots,\bm\xi_{p+1}]\}$, where $s[\bm\xi_{i_1},\cdots,\bm\xi_{i_{p+1}}]$ is the $p$-dimensional simplex with vertices $\{\bm\xi_{i_1},\ldots,\bm\xi_{i_{p+1}}\}$. The empirical version of  $D_F({\bf z})$ is 
$D_{{\hat F}}({\bf z})=  \sum {\bf 1}_{\{{\bf z}\in s[\bm\xi_{i_1},\ldots,\bm\xi_{i_{p+1}}]\}}\big/ {m \choose {p+1}}$). In $\real^2$, 
$D_{\hat F}({\bf z})=
 \sum_{i,l, k} {\bf 1}_{\{{\bf z} \in \Delta(\bm\xi_i, \bm\xi_l, \bm\xi_k)\}}\big/ {{m\choose 3}}$, the fraction of the triangles $\Delta(\bm\xi_i, \bm\xi_l, \bm\xi_k)$ generated from the sample that contains $\bf z$ inside.
Clearly, a point with a larger depth value indicates that it lies more central within the data cloud or closer to the center of the distribution.

By computing the depth values for all data points $\bm\xi_i$'s and then ordering $\bm\xi_i$'s by their descending depth values, we can  obtain the depth order statistics  $\{\bm\xi_{[1]},\ldots, \bm\xi_{[m]}\}$ with an ordering from the deepest (or most central) point $\bm\xi_{[1]}$ to the most outlying $\bm\xi_{[m]}$.
This center-outward ordering naturally gives rise to nested central regions expanding with increasing levels of probability coverage. The convex region spanning the deepest $(1-\alpha)  n$ sample points is referred to as the {\bf $(1-\alpha)$-central region}. Formally, the population and empirical versions of $(1-\alpha)$-central region can be expressed respectively as
\begin{equation}
  \label{eq:DepthRegion}
A_{(1-\alpha); F}= \{{\bf z}: C_F({\bf z}| D_F)\ge  \alpha\} \,\, \hbox{and} \,\, A_{(1-\alpha); {\hat F}}= \{{\bf z}: C_{\hat F}({\bf z}| D_{\hat F})\ge \alpha\}, \quad 0 < \alpha < 1.
\end{equation}
Here, $C_F({\bf z} | D)$ and $C_{\hat F}({\bf z}|D_{\hat F})$ are referred to as {\bf centrality functions} with, respectively,
\begin{equation}
  \label{eq:centrality}
C_F({\bf z} | D) = P_F\{{\bm\xi}: D_F({\bm\xi}) \leq D_F({\bf z})\} \,\,\, \hbox{and} \,\,\, C_{\hat F}({\bf z}|D_{\hat F}) = \frac{1}{m}\sum_{i=1}^{m}{\bf 1}_{\{D_{\hat F}(\bm\xi_i)< D_{\hat F}(\bf z)\}}.
\end{equation}
The central regions $A_{(1-\alpha); {\hat F}}$ are data driven and nonparametric, and are shown to be particularly useful for inference under asymmetric underlying distributions or non-standard asymptotics.

Lemma~\ref{lemma:centrality} below shows important properties of centrality functions. Its proof is in Appendix.

\begin{lemma}
  \label{lemma:centrality}
   Let $\bm\eta$ be a random vector following  a $p$-dimensional distribution $F$. The centrality function in (\ref{eq:centrality}) satisfies these properties:\\
  (a) (Uniform transformation) The transformed variable $C_F(\bm\eta|D_F)$ satisfies $C_F(\bm\eta|D_F)\sim U(0,1)$,
  provided that the depth contours $\{\bm\eta: D_F(\bm\eta)=t\}$ all have probability zero w.r.t. $F$ for any $t>0$. \\
 (b) (Affine-invariance) Let $A$ be a $p\times p$ nonsingular matrix and ${\bf b}$ a $p\times 1$ constant vector. If  both $F(\cdot)$ and $D_F(\cdot)$ are affine invariant, i.e., for any point  ${\bf z} \in \real^p$,  $\tilde F(A{\bf z} +{\bf b}) = F({\bf z})$ and $D_F({\bf z})=D_{\tilde F}(A{\bf z}+{\bf b})$, 
then so is the centrality function $C_F(\cdot)$, i.e., $C_F({\bf z}|D_F)=C_{\tilde F}(A{\bf z}+ {\bf b}|D_{\tilde F})$. 
\end{lemma}
 
Typically, depth functions have been used to rank {\it sample points} and provide a center-outward ordering of sample points in the {\it sample space}, as reviewed above. In this paper, a depth function will be used instead to rank {\it parameter values} and provide an ordering of all parameter values in the {\it parameter space}. Specifically, instead of applying depth ordering to the sample $\bm\xi_i$'s drawn from the distribution $F (\cdot)$, we apply it to the sample CD-random variables $\btheta_i^* $'s drawn from the confidence distribution $H_n(\cdot)$. This center-outward ordering in the parameter space can be interpreted as the plausibility of each parameter value relative to the others. This line of interpretation underlies the proposed CD fusion learning framework and justifies the resulting inferences, e.g.,  using the central regions formed by $\btheta^*$'s as confidence regions for the parameter of interest $\btheta$. This is elaborated further in Sections~\ref{sec:definition} - 5.

\subsection{{\it depth-CD} and {\it depth-CV}: an omnibus form of confidence regions}
\label{sec:definition}

The definition of a CD as a sample-dependent distribution function on the parameter space 
that can represent confidence regions for all possible confidence levels applies to a scalar parameter (as seen in Section 3.1.2) as well as a vector parameter.  
However, mathematical rigor for multi-dimensional CDs has so far been elusive, 
since the region created by the inversion of a multivariate cumulative distribution function may not be unique or suitable for providing any natural form of inference. 
To this end,  \citep{singh2007confidence,xie2013confidence,schweder2016confidence} have proposed to limit confidence regions within a certain subclass. In this paper, we propose to consider the set of center-outward nested confidence regions derived from using data depth, which we refer to as {\it depth-CDs}. The {\it depth-CD}s provide a natural extension of the CD concept in scalar setting to the multiparameter setting.

As discussed in Section~\ref{sec:CD}, a confidence curve (CV), as  plotted in Figure~\ref{fig:CD-CV-Univariate}(b), can provide two-sided confidence intervals for a scalar parameter of all levels, with the intervals expanding outward to two tails as the level of confidence increases. 
The CV in Figure~\ref{fig:CD-CV-Univariate}(b)) clearly ranks parameter values in the parameter space from the center outward as the level $\alpha$ decreases.  In fact, the CV defined in (\ref{eq:CV}) can be re-expressed, using data depth and its associated centrality function, as
 \begin{equation}
 \label{eq:CV1}
C_{H_n}(\theta | D_{\small HS}) = P_{H_n}\{{\xi}: D_{\small HS}({\xi}) \leq D_{\small HS}({\theta})\} = 2 \min\{H_n(\theta), 1-H_n(\theta)\} = CV_n(\theta),
\end{equation}
where $D_{\small HS} (\bm\vartheta)= \inf_E \{P_{H_n}(E): E\ \text{is a closed half-space in}\  \mathbb{R}^p \ and\ \bm\vartheta \in E\}$ is the half-space depth when $p = 1$ and $P_{H_n}$ is the probability measure corresponding to the CD $H_n(\cdot)$ on the parameter space, i.e.,  $P_{H_n}\{( - \infty, t]\} = H_n(t)$.

By extending (\ref{eq:CV1}), we can directly define a CV for a parameter vector $\vartheta \in \bm \Theta \subset \mathbb{R}^p$ as
\begin{equation}
 \label{eq:CV-general}
CV_n(\vartheta) =: C_{H_n}(\vartheta | D) = P_{H_n}\{{\bm \xi}: D({\bm\xi}) \leq D({\vartheta})\},
\end{equation}
where $D$ is a depth function with the associated probability measure $P_{H_n}$ from a multivariate {\it depth-CD} $H_n(\cdot)$ on  $\bm \Theta$. Formally, we define multivariate depth- CD and CV  as follows:
\begin{definition}[{\it depth-CD} and Depth CV]
(A)  A function $H_n(\cdot)\equiv H_n(\bm X_n,\cdot)$ on $\bm\Theta \subseteq \mathbb{R}^p$ is called a depth confidence distribution ({\it depth-CD}) associated with depth function $D$ for a vector-valued parameter $\bm\theta$, if
  (i) it is a distribution function on the parameter space $\bm\Theta$ for any fixed sample set  $\bm X_n$; and
  (ii) the $(1-\alpha)$ ``central region'' of the distribution $H_n(\cdot)$,
$\mathcal{R}_{1-\alpha}(H)=\{\bm\vartheta \in \bm\Theta: C_{H_n}(\bm\vartheta) \geq \alpha \}$,
 is a confidence region for $\bm\theta$ with a coverage probability of $(1-\alpha)$.
Here,  the centrality function associated with depth $D$ and CD $H_n(\cdot)$, i.e., $C_{H_n}(\bm\vartheta)$, is also referred to as depth confidence curve ({\it depth-CV}).  \\
\phantom{aa} If the statements in (ii) holds only asymptotically, then we refer to $H_n$ and $C_{H_n}$ as asymptotic {\it depth-CD} and asymptotic {\it depth-CV}, respectively.
  \label{def:Depth-CD}
\end{definition}

Continuing with half-space depth in the scalar setting, we see that the result in Lemma~\ref{lemma:centrality}(a) resembles $2 \min\{G(Z), 1-G(Z)\}$ for a univariate random variable $Z$ with its cumulative distribution function $G$. 
This result ensures that $\{\theta: CV_n(\theta) \geq \alpha \}  = [H_n^{-1}(a), H_n^{-1}(b)]$, with $a = \min(\alpha/2, 1-\alpha/2 )$ and $b = \max(\alpha/2, 1-\alpha/2)$, is a $(1 - \alpha)$ confidence interval $\theta$.
Similarly, by Lemma~\ref{lemma:centrality}(a),  the $(1-\alpha)$ ``central region'' of {\it depth-CD} $H_n(\cdot)$ or {\it depth-CV} $CV_n(\cdot)$
\begin{equation}
\label{eq:CR}
\mathcal{R}_{1-\alpha}(H_n)=\{\bm\vartheta \in \bm\Theta: C_{H_n}(\bm\vartheta) \geq \alpha \}
\end{equation}
leads to a $(1 - \alpha)$ confidence region for a multivariate $\bm \theta$.
In conclusion, Lemma~\ref{lemma:centrality} ensures that the {\it depth-CD} and {\it depth-CV} (defined in Definition~\ref{def:Depth-CD}) can provide valid center-outward confidence regions of all levels.

We use the familiar bivariate normal to illustrate the above framework of {\it depth-CD} inference.
\begin{example}[Bivariate normal distribution]
  \label{exp:bi-norm}
 Given a random sample $\{\bm Y_i\}_{i=1}^n$ from a bivariate normal distribution $BN(\bm\theta, \Sigma)$, we consider making inference for the mean parameter $\bm\theta$. Let $\bar{\bm Y}_n=\sum_{i=1}^n \bm Y_i/n$. Assuming that $\Sigma$ is known, then the bivariate normal distribution $BN(\bar{\bm Y}_n, \Sigma)$ is a {\it depth-CD} for $\bm\theta$, since: I) $BN(\bar{\bm Y}_n, \Sigma)$ is a sample-dependent distribution function of the parameter space of $\bm \theta$, and  II) the depth contours of $BN(\bar{\bm Y}_n, \Sigma)$, using any depth mentioned in Section 3.1.2, provide valid center-outward confidence regions of all levels.

For a given simulated sample of size $\{\bm Y_i\}_{i=1}^{20}$ under  $\bm\theta=(1,1)$, and $\Sigma=\left(\begin{smallmatrix}1 & 1\\1& 2 \end{smallmatrix}\right)$, using Mahalanobis depth, we obtain the {\it depth-CD} on the parameter space $\bm\Theta$ as a 3D-surface plot in Figure~\ref{fig:CD-Bivariate-Normal}(a). Projecting this 3D plot to $\bm\Theta$ (the two-dimensional plane below) gives depth contours in a grey-color heat map in Figure~\ref{fig:CD-Bivariate-Normal}(b), where the brighter the region, the larger the depth value. A depth contour in Figure~\ref{fig:CD-Bivariate-Normal}(b) connects the points in $\bm\Theta$ which have the same depth value $D_H(\cdot)$.
Corresponding to Figure~\ref{fig:CD-Bivariate-Normal}(b), a similar projection of the 3D-surface plot of the {\it depth-CV} results in Figure~\ref{fig:CD-Bivariate-Normal}(c) showing the contours which connect the points in $\bm\Theta$ with the same centrality value $C_{H_n}(\cdot | D)$. For instance, the peak of {\it depth-CV} corresponds to the deepest (or most central) point in Figure~\ref{fig:CD-Bivariate-Normal}(c), which also corresponds to the highest point in the {\it depth-CD} in Figure~\ref{fig:CD-CV-Univariate}(a) as well as the deepest point in Figure~\ref{fig:CD-CV-Univariate}(b).The {\it depth-CV} in Figure~\ref{fig:CD-Bivariate-Normal}(c) ranks the plausibility of each possible value of the bivariate parameter space $\bm\Theta$. For instance, the black round dot being on the contour with centrality value 0.9 implies that this particular parameter value is deeper than 90\% of all the possible parameter values w.r.t. the confidence distribution $H(\cdot)$ or more plausible than 90\% of all the possible parameter values in $\bm\Theta$.

Inferences about $\bm\theta$ can be derived from the depth CD or {\it depth-CV} with its contours in Figure~\ref{fig:CD-Bivariate-Normal}. For example, the largest elliptical region within the contour of centrality value .1 (labeled with a solid triangle) in Figure~\ref{fig:CD-Bivariate-Normal}(c) is a 90\% confidence region for $\bm\theta$.
The deepest point in all three plots $(0.94,0.92)$, marked by a cross, can be considered the most plausible parameter value and thus a suitable point estimate for $\bm\theta$. This point estimate is shown to be consistent later in Section 3.4.

When $\Sigma$ is unknown, the $BN(\bar{\bm Y}_n, \hat{\Sigma})$ can be shown to be a {\it depth-CD} for $\bm\theta$ asymptotically. Here $\hat{\Sigma}$ is the sample covariance matrix. Similar illustration plots and asymptotic inferences can be drawn accordingly. 
\end{example}

\subsection{{\it depth-CD} and {\it depth-CV}: an omnibus form of $p$-values}
\label{sec:p-value}

To show how {\it depth-CD} and {\it depth-CV} can give rise to an omnibus form of $p$-values, we 
first justify that\Minge{, for a given $\bm\vartheta \in \bm \Theta$, the depth CV} $C_{H_n(\bm X_n,\cdot)}(\bm\vartheta)$ is a {\it limiting p-value} for testing the hypothesis $\Omega_0: \bm\theta=\bm\vartheta$ versus $\Omega_1: \bm\theta \neq \bm\vartheta$.
\cite{liu1997notions} defines a sequence of statistics $p_n$ to be a {\it limiting $p$-value} if $p_n \in [0,1]$ and $p_n$ satisfies
\begin{itemize}[itemsep=-4pt]
  \item[(a)] $\limsup_{n \to \infty} \pr_F\{p_n \leq t\} \leq t$ for all $F \in \Omega_0$ and any $t \in [0,1]$; and
  \item[(b)] $p_n \to 0$ in probability for all $F \in \Omega_1$, as $n \to \infty$.
\end{itemize}
To see why $C_{H_n(\bm X_n,\cdot)}(\bm\vartheta)$ is a limiting $p$-value, we need the simple but useful result below:

\begin{proposition}
  The statement that $\mathcal{R}_{1-\alpha}(H_n(\bm X_n,\cdot))$ is a confidence region for $\bm\theta$ with a coverage probability of $(1-\alpha)$ (Requirement (ii) in Definition~\ref{def:Depth-CD}) is equivalent to the statement that $C_{H_n}(\bm\theta^o)\equiv C_{H_n(\bm X_n,\cdot)}(\bm\theta^o)$, as a function of the sample $\bm X_n$, follows the uniform distribution on [0,1], where $\bm\theta^o$ is the true value of $\bm\theta$.
  \label{prop:equiv}
\end{proposition}

In view of Proposition~\ref{prop:equiv}, $C_{H_n(\bm X_n,\cdot)}(\bm\vartheta)$  is a limiting $p$-value as long as $C_{H_n(\bm X_n,\cdot)}(\bm\vartheta) \to 0$ in probability for all $\bm\vartheta \neq \bm\theta^o$, which is a mild condition that holds generally. Such a CD $p$-value and the classical $p$-value share a similar idea in their approaches, as they both try to assess the degree of inconsistency between the given data and the target null hypothesis by comparing a fixed {\it value} w.r.t. a {\it reference distribution}. But they are fundamentally different, since
\begin{itemize}[itemsep=-4pt]
  \item A classical $p$-value is derived by comparing the observed value $\bm t$ of a statistic $\bm T$ with a reference distribution over the {\it sample space}, i.e., the null distribution of $\bm T$, say $F_{T,0}$;
   \item A CD $p$-value is derived by comparing a hypothesized value $\bm\vartheta$ of the parameter $\bm\theta$ with a reference distribution over the {\it parameter space}, i.e., the {\it depth-CD} $H_n(\bm X_n,\cdot)$.
\end{itemize}
The difference lies in that the assessment of statistical significance, namely, measuring the outlyingness of the value ($\bm t$ or $\bm\vartheta$) w.r.t. the reference distribution ($F_{T,0}$ or $H_n(\bm X_n,\cdot)$), is performed in different (sample or parameter) spaces.

{\bf The CD $p$-value has several advantages}, including:\\
{\bf A1)} Given an inference procedure, the reference distribution $H_n(\bm X_n,\cdot)$ is determined solely  by the sample $\bm X_n$ and it does not depend on the specified value  in the null hypothesis. This is different from the classical $p$-value method where the reference distribution must satisfy the null constraint and thus it may vary depending on the null values.\\
{\bf A2)} Since  the CD method does not need to rely on a test statistic, the CD $p$-value essentially serves simultaneously as a test statistic and a $p$-value. Thus it compress the usual three-step test procedure in the classical $p$-value approach into just one, bypassing: i) construct an explicit test statistic, and then ii) establish or approximate its sampling distribution. This point will be elaborated further in Section~\ref{sec:CD-bootstrap} where bootstrap is used to devise {\it depth-CD} functions.\\
{\bf A3)} The CD reference distribution $H_n(\bm X_n,\cdot)$ carries infinitely many $p$-values $\{C_{H_n(\bm X_n,\cdot)}(\bm\vartheta): \bm\vartheta \in \bm\Theta\}$ for a set of hypothesis testing problems $\{\{\Omega_0: \bm\theta=\bm\vartheta$ versus $\Omega_1: \bm\theta \neq \bm\vartheta\}: \bm\vartheta \in \bm\Theta\}$. This implies that  $C_{H_n}(\cdot)$ provides a distribution of $p$-values over $\bm\Theta$; see for example Figure~\ref{fig:CD-Bivariate-Normal}(c), where the contours of centrality values can be used as $p$-value contours. As a $p$-value in  testing $\bm\theta=\bm\vartheta$ is generally viewed as the strength of evidence from the data in support of the assumption $\bm\theta=\bm\vartheta$, $C_{H_n}(\bm\vartheta)$ can be viewed as a measure of the plausibility of assuming $\bm\theta=\bm\vartheta$. The smaller the value of $C_{H_n}(\bm\vartheta)$, the less plausible $\bm\theta=\bm\vartheta$. In Figure~\ref{fig:CD-Bivariate-Normal}(c), for example, the parameter value marked by the solid triangle is much less plausible than the one marked by the solid round dot. 
To sum up,  a {\it depth-CD} provides a simple but comprehensive summary of data evidence in the sense that a {\it single} reference distribution $H_n(\bm X_n,\cdot)$ can express the plausibility of each $\bm\theta$ value for the entire parameter space $\bm\Theta$. In contrast, the reference distribution $F_{T,0}$ used in the classical $p$-value method expresses only the plausibility of the specific parameter value under the null hypothesis. For instance, it does not simultaneously provide the plausibility of $\bm\theta$ values in the alternative parameter space.

Following along Example~\ref{exp:bi-norm}, for a given sample there, the centrality value $C_{H_n}(\bm\theta_0)$ in the {\it depth-CV} as in Figure~\ref{fig:CD-Bivariate-Normal}(c) would be a $p$-value  for testing 
$\Omega_0: \bm\theta=\bm\theta_0$ versus $\Omega_1: \bm\theta \neq \bm\theta_0$.

\subsection{{\it depth-CD}: its deepest point as a consistent estimator}
\label{depth-CD-point}
Given a data set $\bm X_n$ and a  {\it depth-CD} $H_n(\bm X_n,\cdot)$, we propose to use the deepest point of the {\it depth-CD} $H_n(\bm X_n,\cdot)$ or equivalently the maximum point of the centrality function $C_{H_n(\bm X_n,\cdot)}(\cdot)$, denoted by $\hat{\bm\theta}_n^{MCE}$, as a point estimate for the parameter of interest $\bm\theta$. That is, 
\begin{equation}
  \label{eq:point-est}
  \hat{\bm\theta}_n^{MCE}=\operatorname*{arg\,max}_{\bm\theta \in \bm\Theta} C_{H_n(\bm X_n,\cdot)}(\bm\theta).
\end{equation}
This estimate is referred to as a  {\it maximum centrality estimate} (MCE). Note that, in $\real^1$, the MCE corresponds to the highest value of CV  in Figure~\ref{fig:CD-CV-Univariate}(b) or, equivalently, the median (also central most point) of the CD in Figure~\ref{fig:CD-CV-Univariate}(a). The estimate $\hat{\bm\theta}_n^{MCE}$ extends to general multiparameter settings the idea of using the 'median' or deepest point of a CD function for point estimation. We show below that $ \hat{\bm\theta}_n^{MCE}$ is a consistent estimator under some mild conditions.

\begin{proposition}
   \label{prop:point-est}
  Assume that for any $\epsilon>0$, as $n \to \infty$,
  \[\Delta_n(\epsilon)=\max_{\bm\vartheta_i,\bm\vartheta_j \in\{\bm\vartheta: C_{H_n(\bm X_n,\cdot)}(\bm\vartheta)=\epsilon \}} ||\bm\vartheta_i-\bm\vartheta_j|| \to 0\] in probability. Then, $\hat{\bm\theta}_n^{MCE} \to \bm\theta^o$ in probability. Furthermore, if $\Delta_n(\epsilon)=O_p(a_n)$ for a non-negative sequence $a_n \to 0$, then $\hat{\bm\theta}_n^{MCE} - \bm\theta^o = O_p(a_n)$.
\end{proposition}

The condition $\Delta_n(\epsilon) \to 0$ basically requires that the depth contours $\{\bm\vartheta: C_{H_n(\bm X_n,\cdot)}(\bm\vartheta)=\epsilon \}$ (e.g., the contours in Figure~\ref{fig:CD-Bivariate-Normal}(c)) shrink to a single point (e.g., the cross in Figure~\ref{fig:CD-Bivariate-Normal}(c)) as the sample size $n\to \infty$. For scalar parameters, $\Delta_n(\epsilon)$ is the distance between the two intersection points of the CV and the horizontal dashed line in Figure~\ref{fig:CD-CV-Univariate}(b). The condition $\Delta_n(\epsilon) \to 0$ means that as information increases ($n\to \infty$), the {\it depth-CD} concentrates onto a shrinking area of the parameter space whose measure decreases to zero. This is a mild condition which holds often in practice. Under this condition, Proposition~\ref{prop:point-est} justifies that the estimate  $\hat{\bm\theta}_n^{MCE}$  converges to the true value $\bm\theta^o$. Thus, we have established Property ($\mathcal{P}$-3) of {\it depth-CD}s stated in {\bf Introduction}.

\section{Construct depth-CDs from nonparametric bootstrap}
\label{sec:CD-bootstrap}

This section provides a concrete approach of using nonparametric bootstrap to construct a {\it depth-CD} and derive  inferences for the target parameter vector in an individual study. Given the sample $\{X_1,X_2,\ldots,X_n\}$, assume that  $\hat{\bm\theta}_{n}$  is an estimate of the target parameter $\bm\theta$, where
\begin{equation}
    \hat{\bm\theta}_{n} \equiv \hat{\bm\theta}_{n}(X_{1},X_{2},\ldots,X_{n}).
    \label{eq:theta-est}
\end{equation}
Let $\hat{\bm\theta}_{n}^*=\hat{\bm\theta}_{n}^*(X_{1}^*,X_{2}^*,\ldots,X_{n}^*)$ be a bootstrap estimate of $\bm\theta$, where $\{X_{1}^*,X_{2}^*,\ldots,$ $X_{n}^*\}$ is a bootstrap sample drawn independently from $\{X_{1},X_{2},\ldots,X_{n}\}$ with replacement. Let $B_n$ and $B_n^*$ denote the sampling distribution of $\hat{\bm\theta}_n$ and $\hat{\bm\theta}_n^*$, respectively. Theorem~\ref{thm:CD-bootstrap} below shows that the bootstrap distribution $B_n^*$ is a {\it depth-CD} for $\bm\theta$ asymptotically under the following regularity conditions:

\smallskip
\noindent
{\it (C1) Let $L_n$ be the distribution of $a_n(\hat{\bm\theta}_n-\bm\theta)$ for some positive sequence $a_n \to \infty$, as $n \to \infty$. Assume that $L_n$ converges $D$-regularly to a distribution $L$ in the sense that (i) $L_n$ converges weakly to $L$ as $n \to \infty$; and (ii) $\lim_{n \to \infty}\sup_{x \in \mathbb{R}^p}\mid D(L_n, x)-D(L, x) \mid~ =0$.\\
(C2)  Let $L_n^*$ be the distribution of $a_n(\hat{\bm\theta}_n^*-\hat{\bm\theta}_n)$. Assume that $L_n^*$ converges $D$ regularly to the distribution $L$ almost surely.\\
(C3) The distribution $L$ is continuous and symmetric around 0.\\
(C4) The distribution of $D(L, l)$ is continuous, where the random variable $l \sim L$.
}

\begin{theorem}
  Under the regularity conditions (C1)-(C4), the distribution of $\hat{\bm\theta}_n^*$, conditional on the sample $\{X_{1},X_{2},\ldots,X_{n}\}$, is a {\it depth-CD} for the parameter $\bm\theta$ asymptotically as $n \to \infty$.
  \label{thm:CD-bootstrap}
\end{theorem}

Theorem~\ref{thm:CD-bootstrap} shows that the bootstrap distribution  $B_n^*$ is a {\it depth-CD}, and hence justifies the validity of using $B_n^*$ to make inferences about the parameter vector $\bm\theta$. For example, the deepest point of the distribution $B_n^*$ can be used as a point estimate of $\bm\theta$,  and the central region  $\mathcal{R}_{1-\alpha}(B_n^*)$, as defined in (\ref{eq:CR}), as a $(1-\alpha)$ confidence region for $\bm\theta$. Moreover,  for testing the hypothesis $\bm\theta=\bm\vartheta$ versus $\bm\theta \neq \bm\vartheta$, the value of the centrality function at $\bm\vartheta$, i.e.,  $C_{B_n^*}(\bm\vartheta)$, can be used as a $p$-value. 

This $p$-value approach for hypothesis testing is fundamentally different from traditional approaches, as mentioned in Section~\ref{sec:p-value}. First, the reference distribution here is {\it depth-CD} $B_n^*$, which is fully determined by the sample. Once the sample is given, it does not vary, unlike the traditional approaches. Second, {\it depth-CD} $B_n^*$ is a single reference distribution and provides a $p$-value for testing each parameter value in the entire parameter space $\bm\Theta$. Third, in the derivation of the $p$-value, $C_{B_n^*}(\bm\vartheta)$ does not rely on any test statistic. Essentially, $C_{B_n^*}(\bm\vartheta)$ now simultaneously serves as a test statistic and as a $p$-value. This actually compresses into a single step, namely calculating the centrality of $\bm\vartheta$ w.r.t. {\it depth-CD} $B_n^*$, the usual three steps in traditional testing  procedures, namely identifying a test statistic, establishing its sampling distribution, and then calculating the $p$-value. Fourth, the reference distribution {\it depth-CD} $B_n^*$ is obtained by resampling directly from the empirical distribution, rather than from the null distribution that is usually restricted by parametric assumptions. This also explains why our CD inference here can be obtained without distributional assumptions of the sample.

The idea of connecting bootstrap to data depth for multiparameter inference is not new. For example, it has been used in \cite{liu1997notions}  for hypothesis testing and in \cite{yeh1997balanced} for deriving confidence regions. These two inference methods can be viewed as special cases in our {\it depth-CD} inference framework, since Theorem~\ref{thm:CD-bootstrap} implies that the bootstrap distribution is a {\it depth-CD}. 

Theorem~\ref{thm:CD-pivotal} below is a direct consequence of Lemma~~\ref{lemma:centrality} and Proposition~\ref{prop:equiv}, and it  provides a procedure for constructing {\it depth-CD}s from pivot statistics. 

\begin{theorem}
  \label{thm:CD-pivotal}
Assume that  $A_n(\bm X_n)$ is a nonsingular matrix such that  $A_n(\bm X_n)(\hat{\bm\theta}_n-\bm\theta)$ follows a distribution $Q_n$ that is free of all unknown parameters. Also assume that $\bm\eta_n$ is a random vector, independent of the sample $\bm X_n$, following the distribution $Q_n$. Then, conditional on $\bm X_n$, the distribution function of $(\hat{\bm\theta}_n-A_n(\bm X_n)^{-1}\bm\eta_n)$ is a {\it depth-CD} for $\bm\theta$,
 under the following conditions\\
  (i) The depth $D$ is affine-invariant; and\\
  (ii) The depth contours $\{\bm\eta_n: D_{Q_n}(\bm\eta_n)=t\}$ all have probability zero w.r.t. $Q_n$.
\end{theorem}

Theorem~\ref{thm:CD-pivotal} shows that when the statistic $A_n(\bm X_n)(\hat{\bm\theta}_n-\bm\theta)$ is a pivot, a {\it depth-CD} can be easily derived using the inverse probability function. Returning to Example~\ref{exp:bi-norm} where we  make inference for the mean parameter of a bivariate normal distribution. In this case, we know that $\Sigma^{-1/2}(\bar{\bm Y}_n-\bm\theta)$ is a pivot following a bivariate standard normal distribution and that the three depths mentioned in Section~\ref{sec:definition} are affine-invariant. Thus, by Theorem~\ref{thm:CD-pivotal}, the bivariate normal distribution $BN(\bar{\bm Y}_n, \Sigma)$ is a {\it depth-CD} for $\bm\theta$. When $\Sigma$ is not known, $BN(\bar{\bm Y}_n, \hat{\Sigma})$ is a {\it depth-CD} for $\bm\theta$ asymptotically. In this example, the distribution of the pivot $\Sigma^{-1/2}(\bar{\bm Y}_n-\bm\theta)$, namely $Q_n$ in Theorem~\ref{thm:CD-pivotal}, is structured under certain distributional  assumptions, i.e., $Q_n$ is bivariate standard normal. But generally, as long as $A_n(\bm X_n)(\hat{\bm\theta}_n-\bm\theta)$ can be structured to have (approximately) a parameter-free distribution, Theorem~\ref{thm:CD-pivotal}  can be applied to construct {\it depth-CD}s and draw all forms of inference accordingly, as seen in Section~\ref{sec:depth-CD-inference}.

With the distribution $Q_n$ as a prerequisite,Theorem~\ref{thm:CD-pivotal} may be perceived as applicable only for deriving {\it depth-CD}s in the setting of parametric inference, but its precise formulation can actually shed light on the bootstrap approach in Theorem~\ref{thm:CD-bootstrap} and other general nonparametric approaches for deriving {\it depth-CD}s.
Here is how Theorem~\ref{thm:CD-pivotal} explains intuitively why the nonparametric bootstrap distribution is indeed a {\it depth-CD}.
To avoid making assumptions about the distribution $Q_n$ of the statistic $A_n(\bm X_n)(\hat{\bm\theta}_n-\bm\theta)$, a natural choice is to use the bootstrap distribution of $A_n(\bm X_n)(\hat{\bm\theta}_n^*-\hat{\bm\theta}_n)$ to approximate $Q_n$, i.e., set $\bm\eta_n=A_n(\bm X_n)(\hat{\bm\theta}_n^*-\hat{\bm\theta}_n)$ in Theorem~\ref{thm:CD-pivotal}. Assuming that this approximation is appropriate (e.g., under conditions (C1)-(C4)), Theorem~\ref{thm:CD-pivotal} shows that the distribution of $(2\hat{\bm\theta}_n-\hat{\bm\theta}_n^*)$ $(=\hat{\bm\theta}_n-A_n(\bm X_n)^{-1}A_n(\bm X_n)(\hat{\bm\theta}_n^*-\hat{\bm\theta}_n))$ is a {\it depth-CD} for $\bm\theta$. Note that $(2\hat{\bm\theta}_n-\hat{\bm\theta}_n^*)$ and $\hat{\bm\theta}_n^*$ have an identical distribution (provided that the distribution of $(\hat{\bm\theta}_n^*-\hat{\bm\theta}_n)$ is symmetric, see condition (C3)), since when centering around $\hat{\bm\theta}_n$, the refection image of $(2\hat{\bm\theta}_n-\hat{\bm\theta}_n^*)$ is exactly $\hat{\bm\theta}_n^*$. Hence, the bootstrap distribution of $\hat{\bm\theta}_n^*$ is also a {\it depth-CD}.

\section{Fusion Learning using depth CVs}
\label{sec:CD-fusion} 
\subsection{Combining {\it depth CV}s}
We have shown that the very form of the {\it depth-CD} being an all-encompassing distributional function estimate, rather than a mere point or interval estimate, is the key feature that leads to the omnibus form of all inferences of a parameter. This feature will also be shown to underlie the great flexibility that makes {\it depth-CD}s particularly suited for combining inferences from different and even heterogeneous studies.  

For each individual study, we can obtain a {\it depth-CD} $H_{k,n_k}(\cdot) \equiv H_{k,n_k}(\bm X_k; \cdot)$ and its corresponding {\it depth-CV} $C_{H_{k,n_k}}(\cdot)$ (cf. Definition~\ref{eq:CV-general}) for the parameter $\bm \theta$, $k=1,2,\ldots,K$, from $K$ independent studies. Here we propose a general formula in (\ref{eq:combining-general})
for synthesizing those $K$ individual inference results to draw an overall and efficient inference for the parameter $\bm \theta$,
\begin{equation}
  \label{eq:combining-general}
  C_{(c)}(\bm\theta)= G_c\left(g_c(C_{H_{1,n_1}}(\bm\theta),C_{H_{2,n_2}}(\bm\theta),\ldots,C_{H_{K,n_K}}(\bm\theta))\right).
\end{equation}
Here, $g_c(u_1,u_2,\ldots,u_K)$ is a continuous mapping from $[0,1]^K$ to $\mathbb{R}$ which is increasing in each coordinate, and $G_c(t)\equiv \pr\{g_c(U_1,U_2,\ldots,U_K)\leq t\}$ where $U_k$'s are $i.i.d.$ random variables following U[0,1] distribution. A special yet important case of (\ref{eq:combining-general}) to which we will return often later is 
\begin{equation}
  C_{(c)}(\bm\theta)= F_{(c)} \left\{ w_1 \varphi(C_{H_{1,n_1}}(\bm\theta)) + w_2 \varphi(C_{H_{2,n_2}}(\bm\theta)) + \cdots w_K \varphi(C_{H_{K,n_K}}(\bm\theta)) \right\},
  \label{eq:combining-special}
\end{equation}
with $g_c(u_1,u_2,\ldots,u_K)=w_1 \varphi(u_1) + w_2 \varphi(u_2) + \cdots w_K \varphi(u_K)$, where $\varphi(\cdot)$ is a monotonic increasing  ``transformation function'' and $w_k > 0$ is the weight assigned to the $k$-th study.

Fusion formulas similar to (\ref{eq:combining-general}) and (\ref{eq:combining-special}) have been used in \citep{singh2005combining,xie2011confidence,liu2014exact} to combine CDs for a scalar parameter. However, these do not apply directly to combining {\it depth-CD}s  
for multiparameter inference. If the combining formula (\ref{eq:combining-general}) were applied directly, the resulting function $G_c(g_c(H_{1,n_1}(\bm\theta),H_{2,n_2}(\bm\theta),\dots,H_{K,n_K}(\bm\theta)))$ would not yield any valid statistical inference. To mitigate this shortcoming, our proposal in (\ref{eq:combining-general}) and (\ref{eq:combining-special}) combines the {\it depth-CV}s through their corresponding centrality functions $C_{H_{k,n_k}}(\bm\theta)$'s to obtain $C_{(c)}(\bm\theta)$ (cf. (\ref{eq:combining-general})) and the  desired overall inference.

\begin{theorem}
  \label{thm:fused-CD}
  Given the individual {\it depth-CD}s $H_{k,n_k}(\bm\theta)$, $k=1,\ldots,K$, the following forms of inference for the common parameter $\bm\theta$ derived from the combined \Minge{depth CV} function $C_{(c)}(\bm\theta)$ in (\ref{eq:combining-general}) are valid.\\
  (a) (Hypothesis testing) For testing the null hypothesis \[\Omega_0: \bm\theta=\bm\vartheta \quad \text{versus}\quad \Omega_1: \bm\theta \neq \bm\vartheta,\] $C_{(c)}(\bm\vartheta)$ is a limiting  $p$-value, as discussed in Section~\ref{sec:p-value}, provided that $C_{H_{k,n_k}}(\bm\vartheta)\to 0$ in probability for all $\bm\vartheta \neq \bm\theta^o$. Here $\bm\theta^o$ is the true parameter value.\\
  (b) (Confidence region) A $(1-\alpha)$ confidence region for $\bm\theta$ is
\[
  \mathcal{R}^{(c)}_{1-\alpha}(H_{1,n_1},H_{2,n_2}\ldots,H_{K,n_K})=\{\bm\theta \in \bm\Theta: C_{(c)}(\bm\theta) \geq \alpha \}.
\]
(c) (Point estimation) Assume that $C_{(c)}(\bm\theta)$ achieves its maximum at $\hat{\bm\theta}_{(c)}$, i.e.,
\begin{equation}
  \hat{\bm\theta}_{(c)}=\max_{\bm\theta \in \bm\Theta} C_{(c)}(\bm\theta).
  \label{eq:estimate-com}
\end{equation}
Then, $C_{(c)}(\bm\theta)$ is a consistent estimator for $\bm\theta^o$. Specifically, as $n_1, n_2, \ldots, n_K \to \infty$, $\hat{\bm\theta}_{(c)} \to \bm\theta^o$ in probability, provided that $C_{H_{k,n_k}}(\cdot)$ is continuous and
\[\Delta_{k, n_k}(\epsilon)=\max_{\bm\vartheta_i,\bm\vartheta_j \in\{\bm\vartheta: C_{H_{k,n_k}(\bm X_{k,n_k},\cdot)}(\bm\vartheta)=\epsilon \}} ||\bm\vartheta_i-\bm\vartheta_j|| \to 0\] in probability, for $k=1,2,\ldots,K$.
\end{theorem}


Theorem~\ref{thm:fused-CD} justifies that the overall inferences based on the combined centrality function $C_{(c)}(\bm\theta)$ can be made in  ways similar to those based on centrality functions from individual studies. For example, Theorem~\ref{thm:fused-CD}(a) shows that, similar to each individual centrality function, the combined function $C_{(c)}(\bm\theta)$ is a single invariant (under the given samples) function defined on the parameter space and it provides infinitely many $p$-values for testing all $\bm\theta$ values in the entire parameter space. It expresses the relative ranking or level of plausibility of each $\bm\theta$ value w.r.t. the totality of evidence collected from all studies. This expression of relative ranking of plausibility adapts readily to the common interpretation of a $p$-value. Theorem~\ref{thm:fused-CD}(b) describes a $(1-\alpha)$ confidence region for $\bm\theta$ as the collection of parameter values whose $C_{(c)}(\bm\theta)$ is no less than $\alpha$. Theorem~\ref{thm:fused-CD}(c) shows that the maximizer of  the combined $C_{(c)}(\cdot)$ is a valid point estimator.

Our fusion learning does not rely on parametric assumptions, if  (\ref{eq:combining-general}) or (\ref{eq:combining-special}) is applied to {\it depth-CD}s from nonparametric approaches, such as bootstrap. 
This fusion approach is broadly applicable. It is valid as long as the input functions $H_{k,n_k}(\bm\theta)$'s are {\it depth-CD}s (or asymptotically).

\bigskip
\noindent{\bf $\bullet$ Higher-order accuracy of $C_{(c)}(\cdot)$ and its inference results --}\\
The {\it depth-CD}s obtained by bootstrap are not {\it exact}, in the sense they only satisfy asymptotically the requirement (ii) in Definition~\ref{def:Depth-CD} or (ii)$^\prime$  in Proposition~\ref{prop:equiv}. To see how this approximation accuracy affects the accuracy of the inference results, we consider the example of a univariate common mean problem, where a  CD obtained by the regular bootstrap in Section~\ref{sec:CD-bootstrap} is $H_{k,n_k}(\theta)=\pr^*\{\bar{X}_k^* \leq \theta\}$. Such a CD can yield confidence regions whose coverage probability approximates the nominal value. A better accuracy can be achieved by using the bootstrap $t$ \citep{efron1994introduction}. This bootstrap method generates a second-order accurate CD
$
  H_{k,n_k}^{(t)}(\theta)=\pr^*\left\{(\bar{X}_k^*-\bar{X}_k)/S^* \geq \bar{X}_k-\theta/S \right\},
$
where $S$ is an estimate of the standard deviation. It would be interesting to know whether or not the improved accuracy in the input CDs is carried over to the combined outcome.  
It is worth noting that our fusion approach in (\ref{eq:combining-special}) generally does preserve the order of accuracy of the individual {\it depth-CD}s, even if they are not {\it exact}. To state the result, we first define the order of accuracy for a {\it depth-CD}.
\begin{quote}
{\it A depth-CD function $H_n(\cdot)\equiv H_n(\bm X_n,\cdot)$ on $\bm\Theta \subseteq \mathbb{R}^p$ is said to be $j$th-order accurate, if the random variable  $C_{H_n}(\bm\theta^o)\equiv C_{H_n(X_n,\cdot)}(\bm\theta^o)$, where $\bm\theta^o$ is the true value of $\bm\theta$, converges in distribution to the uniform distribution on (0,1) at the order of $n^{-j/2}$, i.e., $\pr\left\{C_{H_n}(\bm\theta^o) \leq a \right\} - a =  O(n^{-j/2})$ for any $a \in (0,1)$. If a {\it depth-CD} function $H_n(\cdot)$ is $j$-th order accurate, the coverage probability of $\mathcal{R}_{1-\alpha}(H_n)$ in (\ref{eq:CR}) converges to 
its nominal level at the rate of $O(n^{-j/2})$.}
\end{quote}
\begin{theorem}
 \label{thm:accuracy}
(Accuracy of $C_{(c)}(\cdot))$. For the $k$-th study ($k=1,2,\ldots,K$), assume that $n_k/n$ converges to a constant $a_k$ and also that  its {\it depth-CD} function $H_{k,n_k}(\bm\theta)$ is $j$th-order accurate uniformly, in the sense that $\pr\left\{C_{H_{k,n_k}}(\bm\theta^o) \leq a \right\} - a =  O(n^{-j/2})$ uniformly for all $a \in (0,1)$ as $n \to \infty$.
Then the combined function $C_{(c)}(\bm\theta)$ in its general form (\ref{eq:combining-general}) is also $j$th-order accurate.
\end{theorem}

Our numerical studies in Section 6 show that, even in small-sample cases, the overall inferences are quite accurate, when the input CD functions $H_{k,n_k}^{(t)}(\bm\theta)$'s are obtained by the bootstrap $t$.

\bigskip
\noindent{\bf $\bullet$ Bahadur efficiency of $C_{(c)}(\cdot)$ -- }\\
The fusion formula in (\ref{eq:combining-general}) provides a general class of fusion approaches for synthesizing nonparametric or parametric inferences. We show here that among this general class, a specific form of (\ref{eq:combining-special}) with $w_k=1$ for all $k$ and $\varphi(t)=\log(t)$ yields the most efficient combination in terms of achieving Bahadur efficiency. Following the ideas in \cite{littell1973asymptotic} and \cite{singh2005combining}, we define the concept of {\it Bahadur slope} for a {\it depth-CV}.
\begin{definition}
  A nonnegative function $S_{\bm\lambda}(b)\equiv S_{\bm\lambda}(b;\bm\theta^o)$ is said to be the Bahadur slope for the {\it depth-CV} function $C_{H_n}(\cdot)$ along the direction $\bm\lambda$, where $\bm\lambda \in \mathbb{R}^p$ and $\|\bm\lambda\|=1$, if $S_{\bm\lambda}(b) \equiv \lim_{n \to \infty} -\log \{C_{H_n}(\bm\theta^o+b \bm\lambda) \}/n$ almost surely for any non-zero $b \in \mathbb{R}$.
\end{definition}

The Bahadur slope $S_{\bm\lambda}(b)$ defined above reflects the rate, in an exponential scale, at which $C_{H_n}(\bm\theta^o+b \bm\lambda)$ decays toward zero as the sample size increases. The larger the slope, the more efficient the {\it depth-CV} in Bahadur's sense. In the multiparameter case where the {\it depth-CD} $H_n(\bm\theta)$ is a multivariate distribution, we need Bahadur slope functions $S_{\bm\lambda}(b)$ along each direction $\bm\lambda$ to characterize how fast the tails of the distribution decay to zero.

The Bahadur slope provides a means assessing the efficiency of the proposed fusion method (\ref{eq:combining-general}). Specifically, the theorem below establishes an upper bound of the Bahadur slope (i.e., the fastest possible rate of tail decay) for the combined function $C_{(c)}(\bm\theta)$. It also suggests a specific combination formula for achieving exactly this bound.

\begin{theorem}
  \label{thm:bahadur}
  Under $\bm\theta=\bm\theta^o$ and $n_k=\{a_k+o(1)\}n$, as $n \to \infty$, the following inequality holds for any fused function $C_{(c)}(\bm\theta)$ as defined in the general fusion formula (\ref{eq:combining-general})
  \begin{equation}
    \label{eq:bahadur-bound}
    \limsup_{n \to \infty} - \log \{C_{(c)}(\bm\theta^o+b \bm\lambda) \} /n \leq \sum_{k=1}^K a_k S_{k,\bm\lambda}(b).
  \end{equation}
  Furthermore, let $C_{(c)}^{log}$ denote the fused function in its specific form (\ref{eq:combining-special}) when $w_k=1$ for all $k$ and $\varphi(t)=\log(t)$. Then,
  \begin{equation}
    \label{eq:bahadur-E}
    \lim_{n \to \infty} - \log \{C_{(c)}^{log}(\bm\theta^o+b \bm\lambda) \} /n = \sum_{k=1}^K a_k S_{k,\bm\lambda}(b).
  \end{equation}
\end{theorem}

Theorem~\ref{thm:bahadur} states that the Bahadur slope of any combined function $C_{(c)}(\cdot)$ derived from (\ref{eq:combining-general}) has an upper bound, and that this upper bound can be achieved by taking $w_k=1$ for all $k$ and $\varphi(t)=\log(t)$ in (\ref{eq:combining-special}). In this case, the  explicit formula for combining {\it depth-CV}s is
\begin{equation}
  \label{eq:combining-Fisher}
  C_{(c)}^{log}(\bm \theta)=\pr \left\{\chi_{2K}^2 \geq -2 \sum_{k=1}^K \log(C_{H_{k,n_k}}(\bm\theta))\right\}.
\end{equation}
This formula turns out to be the same as Fisher's method used for combining $p$-values. The optimality of this particular choice does not rely on the direction $\bm\lambda$. Thus, an interesting implication is that if we use (\ref{eq:combining-Fisher}) to combine {\it depth-CV}s, the highest Bahadur slope (or the fastest rate of tail decay) will be achieved along {\it every} direction (i.e., the line spanned by $\bm\theta^o+b \bm\lambda$ as $b$ varies). The optimality established in Theorem~\ref{thm:bahadur} is a global, rather than merely directional,~property~of~(\ref{eq:combining-Fisher}).

\subsection{Fusion of heterogeneous studies}
\label{sec:heter}
Our fusion framework is general and can cover complex and irregular settings containing heterogeneous studies.  
Study heterogeneity arises often in practice, due to different study designs, populations or outcomes, as seen in the applications in \citep{chen2013using,yang2014efficient,liu2015multivariate,chatterjee2016constrained,gao2017data}.
 In the presence of heterogeneous studies, the parameter of interest may not be estimable in  some studies. These studies are often excluded from conventional analyses, which can result in a nonnegligible loss of information. Our fusion method (\ref{eq:combining-general}) can be extended to incorporate heterogenous studies in the analysis. The theoretical results established in previous sections remain valid and applicable as well.

To accommodate heterogenous studies, we give up the assumption that $\bm\theta_k$ is estimable in each study. Instead, we assume only that a certain mapping of $\bm\theta_k$, denoted by $\tilde{\bm\theta}_k(=\bm f_k(\bm\theta_k))$ as in (\ref{eq:assumption-heter}), is estimable 
and its corresponding {\it depth-CD} $H_{k,n_k}(\tilde{\bm\theta}_k)$ for $\tilde{\bm\theta}_k$ can be derived, say, using bootstrap. With a minor modification, the general fusion formula (\ref{eq:combining-general}) is still applicable to combining {\it depth-CD}s from different $\tilde{\bm\theta}_k$'s for making the overall inference about the common parameter of interest $\bm\theta$. More specifically,
\begin{equation}
  \label{eq:combining-general-heter}
  C_{(c)}^{Het}(\bm\theta)= G_c\left(g_c(C_{H_{1,n_1}}(\tilde{\bm\theta}_1),C_{H_{2,n_2}}(\tilde{\bm\theta}_2),\ldots,C_{H_{K,n_K}}(\tilde{\bm\theta}_K))\right).
\end{equation}
Similar to (\ref{eq:combining-special}), a special case is
\begin{equation}
  C_{(c)}^{Het}(\bm\theta)= F_{(c)} \left\{ w_1 \varphi(C_{H_{1,n_1}}(\tilde{\bm\theta}_1)) + w_2 \varphi(C_{H_{2,n_2}}(\tilde{\bm\theta}_2)) + \cdots w_K \varphi(C_{H_{K,n_K}}(\tilde{\bm\theta}_K)) \right\}.
  \label{eq:combining-special-heter}
\end{equation}

Theorem~\ref{thm:fused-CD-heter} below shows how to use $C_{(c)}^{Het}(\bm\theta)$ in (\ref{eq:combining-general-heter}) or (\ref{eq:combining-special-heter}) to make valid combined inference about $\bm\theta$. Here, we require that $\bm\theta$ be identifiable in the combined function $C_{(c)}^{Het}(\bm\theta)$. Following \cite{rothenberg1971identification} and \cite*{little2010parameter}, we say that $\bm\theta$ is (locally) {\it identifiable} if for any $\bm\theta \in \bm\Theta$, there is no $\bm\vartheta \neq \bm\theta$ (in a neighborhood of $\bm\theta$) such that $C_{(c)}^{Het}(\bm X_1, \ldots, \bm X_K; \bm\theta)=C_{(c)}^{Het}(\bm X_1, \ldots, \bm X_K;  \bm\vartheta)$ almost surely.

\begin{theorem}
  \label{thm:fused-CD-heter}
Consider the setup in (\ref{eq:assumption-heter}) and the given {\it depth-CD}s $H_{k,n_k}(\tilde{\bm\theta}_k)$ for $\tilde{\bm\theta}_k$, $k=1,\ldots,K$. Assume that the parameter $\bm\theta$ is identifiable in the combined function $C_{(c)}^{Het}(\bm\theta)$ in its general form (\ref{eq:combining-general-heter}). Then, the following inferences derived from $C_{(c)}^{Het}(\bm\theta)$ are valid. \\
  (a) (Hypothesis testing) For testing the null hypothesis \[\Omega_0: \bm\theta=\bm\vartheta \quad \text{versus}\quad \Omega_1: \bm\theta \neq \bm\vartheta,\] $C_{(c)}^{Het}(\bm\vartheta)$ is a limiting  $p$-value, in the sense as discussed in Section~\ref{sec:p-value}, provided that $C_{H_{k,n_k}}(\bm\vartheta_k)\to 0$ in probability for all $\bm\vartheta_k \neq \bm\theta_k^o$.\\
  (b) (Confidence region) A $(1-\alpha)$ confidence region for $\bm\theta$ is
\[
  \mathcal{R}^{(c)}_{1-\alpha}(H_{1,n_1},H_{2,n_2}\ldots,H_{K,n_K})=\{\bm\theta \in \bm\Theta: C_{(c)}^{Het}(\bm\theta) \geq \alpha \}.
\]
(c) (Point estimation) Assume that $C_{(c)}^{Het}(\bm\theta)$ achieves its maximum at $\hat{\bm\theta}_{(c)}$, i.e.,
\begin{equation*}
  \hat{\bm\theta}_{(c)}=\max_{\bm\theta \in \bm\Theta} C_{(c)}^{Het}(\bm\theta).
\end{equation*}
Then, $C_{(c)}(\bm\theta)$ is a consistent estimator for $\bm\theta^o$. Specifically, as $n_1, n_2, \ldots, n_K \to \infty$, $\hat{\bm\theta}_{(c)} \to \bm\theta^o$ in probability, provided that $C_{H_{k,n_k}}(\cdot)$ is continuous and
\[\Delta_{k, n_k}(\epsilon)=\max_{\bm\vartheta_i,\bm\vartheta_j \in\{\bm\vartheta: C_{H_{k,n_k}(\bm X_{k,n_k},\cdot)}(\bm\vartheta)=\epsilon \}} ||\bm\vartheta_i-\bm\vartheta_j|| \to 0\] in probability, for $k=1,2,\ldots,K$.
\end{theorem}

Theorem~\ref{thm:fused-CD-heter} justifies the validity of using the modified combined {\it depth-CV} to draw overall inferences from heterogenous studies, which is the counterpart of Theorem~\ref{thm:fused-CD} in the case of homogeneous studies. In fact, the counterparts of Theorems~\ref{thm:accuracy}-\ref{thm:bahadur}  can also be established to obtain the same theoretical results of high-order accuracy and Bahadur efficiency under heterogenous studies. In short, the combining in (\ref{eq:combining-general-heter}) preserves the order of accuracy of each individual study, and it achieves Bahadur efficiency when $w_k=1$ for all $k$ and $\phi(t)=\log(t)$.

Compared to the meta-analysis of heterogeneous studies in \cite{liu2015multivariate}, our fusion method here is more general. \cite{liu2015multivariate} requires normality of the distribution of summary statistics. Our fusion method does not require a parametric form of distributional assumptions. If each individual {\it depth-CD} is derived using the nonparametric bootstrap, then the inference drawn from the combined function is also nonparametric. When it is reasonable to make an assumption of the underlying distribution, we can derive {\it depth-CD}s using Theorem~\ref{thm:CD-pivotal} and  our fusion method is still applicable and yields valid inferences.

\section{Simulation studies}
To demonstrate the theoretical advantages of our fusion method, we conduct simulation studies for the common mean problem and meta-analysis of correlation coefficients.

\subsection{The common mean problems}

Making inference on the common mean parameter of multiple populations is referred to as the common mean problem. This problem has been investigated extensively, see, e.g., \citealp{lin2007generalized,pal2007revisit}, and the references therein. 
Traditional approaches rely on the assumption that the sample of each study is drawn from a normal distribution. The normality assumption however is often unrealistic in practice, and it can be hardly justified when the sample size is small. To the best of our knowledge, there have not been any systematic investigation of the common mean problem in general and non-normal situations. 

Our framework of fusion learning readily applies to the common mean problem, in both  normal and non-normal settings. In this section, we examine its numerical performance, in comparison with that of several existing methods associated with the well-known Graybill-Deal estimator \citep{graybill1959combining}. The numerical results show that without the normality assumption, our fusion method has the following advantages: 1) it preserves inference accuracy in hypothesis testing/confidence regions; 2) its point estimator has less bias and is more efficient; and 3) it achieves a gain of efficiency in the presence of heterogeneous studies.

In the multiparameter setting, the Graybill-Deal estimator is
\[
 \hat{\bm\mu}_{GD}=\left\{\sum_{k=1}^K n_k \bm S_k^{-1} \right\}^{-1} \sum_{k=1}^K n_k \bm S_k^{-1} \bar{\bm X}_k,
\]
where $\bar{\bm X}_k=\frac{1}{n_k}\sum_{j=1}^{n_k}\bm X_{k,j}$ and $\bm S_k=\frac{1}{n_k-1}\sum_{j=1}^{n_k} (\bm X_{k,j}-\bar{\bm X}_k)(\bm X_{k,j}-\bar{\bm X}_k)^\prime$. This estimator yields confidence intervals and $p$-values by considering the statistic \citep{lin2007generalized}
\begin{equation}
  \label{eq:test-stat-JK}
  \sum_{k=1}^K w_k T^2_k = \sum_{k=1}^K w_k n_k(\bar{\bm X}_k-\bm\mu_0)^\prime \bm S_k^{-1}(\bar{\bm X}_k-\bm\mu_0).
\end{equation} 
Assume that $\bm X_{k,j}$ follows a multivariate normal distribution, then $T^2_k$'s are Hotelling's $T^2$ statistics and $\frac{n_k-p}{p(n_k-1)}T^2_k \sim F_{p,n_k-p}$. Thus, the  statistic in (\ref{eq:test-stat-JK}) follows a weighted convolution of multiple $F$ distributions. We evaluate (\ref{eq:test-stat-JK}) in the construction of confidence regions and hypothesis testing when $w_k \equiv 1$ (referred to as the GD method) and $w_k=Var(T^2_k)^{-1}=\{2p(n_k-1)^2(n_k-2)\}/\{(n_k-p-2)^2(n_k-p-4)\}$ (referred to as the KJ method, \citealp{jordan1995confidence}). If the normality assumption holds, both the GD and KJ methods are exact in the sense that the test (or confidence region) achieves the nominal type I error (or coverage probability), since the exact distribution of (\ref{eq:test-stat-JK}) is known. We also consider a method based on the central limit theorem (CLT). This method needs a weaker assumption, namely that $\bar{\bm X}_k$ only approximately follows a normal distribution. The inference relies on the statistic
\[
\left\{\sum_{k=1}^K n_k \hat{\bm\Sigma}_k^{-1}(\bar{\bm X}_k-\bm\mu_0)\right\}^\prime \left\{\sum_{k=1}^K n_k \hat{\bm\Sigma}_k^{-1} \right\}^{-1} \left\{\sum_{k=1}^K n_k \hat{\bm\Sigma}_k^{-1}(\bar{\bm X}_k-\bm\mu_0)\right\},
\]
where $\hat{\bm\Sigma}_k=(n_k-1)\bm S_k/n_k$. This statistic follows $\chi^2$ distribution with $p$ degrees of freedom.

To implement our nonparametric CD fusion method (\ref{eq:combining-special}), we use half-space depth and Bahadur-efficient combination as in (\ref{eq:combining-Fisher}). The bootstrap $t$ is used, when applicable, with 2000 bootstrap replicates in each run. We compare our method with the GD, JK and CLT methods under the following scenarios. Without loss of generality, we set $K=2$ and consider bivariate distributions.

\noindent{\bf Scenario 1 (Normal distribution)} Let $\bm X=(Z_1,Z_2)^\prime$ follow a bivariate normal distribution with
$\bm \mu_0=E(\bm X)=(0,0)^\prime$, $\sigma(Z_1)=1$, $\sigma(Z_2)=2$, and $Corr(Z_1,Z_2)=\rho$. In Study 1, $\bm X_{1,j}\ i.i.d.\sim~\bm X$ with $\rho=0.8$,
and in Study 2: $\bm X_{2,j}\ i.i.d.\sim~\bm X$ with $\rho=0.3$.

\noindent{\bf Scenario 2 ($\chi^2$ distribution)} Let $\bm X=(Z_1^2,Z_2^2)^\prime$ where $(Z_1,Z_2)^\prime$ follows the same bivariate normal distribution as in Scenario 1. The true value $\bm\mu_0=E(\bm X_{1,j})=E(\bm X_{2,j})=(1,4)^\prime$.

\noindent{\bf Scenario 3 (Cauchy distribution)} Let $\bm X=(Z_1,Z_2)^\prime$ follow a bivariate Cauchy distribution where $Z_1$ and $Z_2$ are independent. The scale parameters $\sigma(Z_1)=1$ and $\sigma(Z_2)=2$ in Study 1, and $\sigma(Z_1)=4$ and  $\sigma(Z_2)=2$ in Study 2. The location parameters
$\bm \mu_0=(0,0)^\prime$ in both studies.

\medskip
\noindent{$\bullet$ \bf Inference accuracy in hypothesis testing/confidence regions}\\
To assess inference accuracy, we present the null distribution of $p$-values in Figures~\ref{fig:CD-combining-normal}-\ref{fig:CD-combining-cauchy} (based on 10000 simulation replications). The deviation of this distribution from the U(0,1) distribution depicts the difference between the actual and nominal type I error rates in hypothesis testing, or equivalently, the difference between the actual and the nominal coverage probabilities of confidence regions. When the sample distribution is normal, Figure~\ref{fig:CD-combining-normal} shows that the null distribution of $p$-values aligns well with the U(0,1) distribution for all the methods considered, except that the CLT method is slightly off the target line. However, when the sample distribution is non-normal, such as $\chi^2$, Figure~\ref{fig:CD-combining-chisq} shows a notable deviation for GD, JK and CLT methods. More details on those deviations can be seen from the empirical values reported in Table~\ref{tab:ecdf-pvalue} for a set of specific points. The numerical values in the table can also be viewed as the (nominal or actual) type I error rates. Boldfaced are the values with a notable deviation from their nominal levels. For example, when the nominal probability (or the type I error rate) is 0.05, the actual probability is 0.18, 0.18, and 0.25, respectively, for GD, JK and CLT methods. Such a substantial deviation indicates a non-negligible loss of inference accuracy and raises serious concerns on using those methods for inference. Only our CD method yields a null distribution following very closely the target distribution. This example shows that our CD method, due to its nonparametric nature, is robust against the violation of the normality assumption.
In Scenario 3 we sample from a bivariate Cauchy distribution, whose mean does not exist, and our inference is on the location parameter instead. Since the moments of Cauchy distributions do not exist, it is not surprising to see in Figure~\ref{fig:CD-combining-cauchy} that GD, JK and CLT methods all exhibit an appreciable loss of inference accuracy. Again, our method remains approximately accurate, when using the median in (\ref{eq:theta-est}) to construct {\it depth-CD}s. The advantage of CD method seen in Figure~\ref{fig:CD-combining-cauchy} is also confirmed numerically in Table~\ref{tab:ecdf-pvalue}, where the actual type I error rates are quite close to the nominal levels. This example highlights the flexibility of our method in adapting easily to irregular situations where moments of the distribution do not exist.

\medskip
\noindent{$\bullet$ \bf Bias and efficiency in point estimation}\\
We compare our CD point estimator in (\ref{eq:estimate-com}) and Graybill-Deal estimator $\hat{\bm\mu}_{GD}$ in estimating the common mean (or location) parameter $\bm\mu=(\mu_1,\mu_2)^\prime$. The distribution of estimates (based on 1000 simulation replications) is presented as boxplots in Figure~\ref{fig:estimate}. When the sample  distribution is normal, it shows in the first column that both estimators 1) are unbiased; and more interestingly, 2) have comparable variabilities. More precisely, the standard errors of GD and CD estimates are 0.125 and 0.126 for $\mu_1$, and 0.258 and 0.261 for $\mu_2$, respectively. 
This observation implies that although the CD method is nonparametric, it sustains negligible efficiency loss compared to the GD method which does make use of the parametric assumption. When the sample distribution is $\chi^2$, the second column of Figure~\ref{fig:estimate} shows that the variabilities of the two estimators are still comparable, but the GD estimator now shows a notable bias, whereas the CD estimator remains unbiased. When the sample distribution is Cauchy, the third column of Figure~\ref{fig:estimate} shows that both estimators are unbiased, but the CD estimator has much smaller variability than the GD estimator, which indicates that the CD method is more efficient. To summarize, in the absence of normality, the CD estimator outperforms the GD estimator in terms of both unbiasedness and efficiency.

\medskip
\noindent{$\bullet$ \bf Gain of efficiency in the presence of heterogeneous studies}\\
We consider a setting of heterogeneous studies by replicating the two studies in Scenario 1 (bivariate normal) and assuming that the two replicated studies are irregular, in that only the sum of the two components of the random vector $\bm X$ is observed. We are interested in combining inferences from all four studies. Here neither of the two marginal means $\mu_1$ and $\mu_2$ is estimable in all studies, but the sum $\mu_1+\mu_2$ is. The GD estimator $\hat{\bm\mu}_{GD}$ can combine only the two regular studies but discard the irregular ones, whereas our CD estimator in (\ref{eq:combining-special-heter}) can incorporate the two irregular studies as well. This same simulation is repeated under Scenario 2 ($\chi^2$ ). To visualize the gain of efficiency in combining the inferences from all four studies, we present in Figure~\ref{fig:estimate-hetero} the boxplots of the GD and CD estimates of $\mu_2$ (based on 1000 simulation replications). 
The boxplots show  that in both normal and non-normal cases, our CD estimator, by combining all studies, is less variable and thus achieves a greater efficiency. Moreover, our CD estimator still remains almost unbiased in the non-normal case. This phenomenon highlights again the flexibility of our fusion method in accommodating a broad class of study heterogeneity.

\subsection{Meta-analysis of correlation coefficients}
In social and behavioral sciences, correlation coefficients, being invariant to the measuring scale, are often used to represent the size of an effect. The meta-analysis of such an effect size has long been used as a tool to draw a more comprehensive conclusion on the bivariate association; see \cite{schulze2004meta} for an in-depth discussion. Classical meta-analysis inference methods for correlations, such as Fisher's z-transformation, assume that the samples of $(X_k,Y_k), k=1,\ldots,K$, all follow bivariate normal distributions. When such an assumption is violated, inference outcomes could be invalid. In what follows, we show that our CD fusion method readily applies to meta-analysis of correlation coefficients, without requiring any parametric assumptions.

To illustrate the CD fusion method, we use the Pearson sample correlation $r$ as an estimate of the correlation coefficient $\rho$ in (\ref{eq:theta-est}), and apply regular bootstrap (with 2000 replicates) to construct a {\it depth-CD} in each study. To combine {\it depth-CD}s, we use half-space depth and Bahadur-efficient combination (\ref{eq:combining-Fisher}). We compare our method with a naive method and the  Hedges-Olkin (HO) method \citep{schulze2004meta}. The naive method merges the data sets as if all the data are from a single source. It then calculates the sample correlation $r$ and applies Fisher's z-transformation $z=\frac{1}{2}\log(\frac{1+r}{1-r})$, where $z$ follows approximately a normal distribution with mean 0 and variance $1/(n-3)$. The HO method obtains Fisher's z-transformed statistic $z_k$ from each study, and combines them using $\bar{z}=\sum_{k=1}^K (n_k-3)z_k/\sum_{k=1}^K (n_k-3)$. The inference is based on that $\bar{z}\sqrt{\sum_{k=1}^K (n_k-3)}$ follows approximately the $N(0,1)$ distribution. Figure~\ref{fig:corr} compares the three methods by examining  the null distribution of $p$-values for testing the hypothesis $H_0: \rho=0$. When the samples of $(X_k,Y_k)$ indeed follow a bivariate normal distribution, the upper row of Figure~\ref{fig:corr} shows that the distribution of each $p$-value approximates the U(0,1) distribution quite well. This observation indicates that all three methods lead to valid inference in normal cases. In the absence of normality, we let $X_k=Z_k$ and $Y_k=Z_k^2$ where $Z_k \sim N(0,1)$. The lower row of Figure~\ref{fig:corr} shows that the $p$-value distributions of the naive and HO methods deviate substantially from the $U(0,1)$ distribution. More specifically, the type I error rates ($\alpha=0.05$) are 0.38 and 0.37, respectively. The results indicate that these two methods may lead to invalid  inference in non-normal cases. The $p$-value distribution of CD method remains very close to the $U(0,1)$ distribution, which is indicative of its robustness to changing distribution assumptions.

\section{Case study: Analysis of aircraft landing performance}
\label{sec:FAA-example}

Recall from the {\bf Introduction} the motivating example from the FAA project on investigating whether or not aircraft landing operations generally comply with the FAA recommendation that the height of the aircraft at the crossing of runway threshold  be around 15.85m  and touchdown distance be around 432m from runway threshold. This question can be addressed by testing 
\begin{equation}
  \label{hyp:FAA}
   H_0: \bm\mu=(15.85,432)^\prime \quad \text{versus}  \quad H_1: \text{otherwise},
\end{equation}
where $\bm\mu$ is the mean vector for the height at the runway threshold and touch down distance.

We are given landing records of two fleets of aircraft, 820  from {\it Airbus} and 1976 from {\it Boeing}. In view of the large samples, an intuitive approach would be just to apply Hotelling's $T^2$ test to the entire sample of 2796 landing records, pooling together both fleets. This yields a $p$-value of 0.942, which would suggest that there is no evidence supporting that the landing performances do not comply with the FAA recommendation. This intuitive approach for combining two studies however is flawed, since it implicitly assumes that the two studies follow the same distribution and thus fails to account for the difference underlying the two studies, shown clearly in Figure~\ref{fig:FAA0}. After all, 
it is realistic to expect difference in performance from different aircraft manufactured by different makers or of designs.

Accommodating such potential study heterogeneity, our fusion learning method can synthesize evidence from the two studies to provide a valid answer to the question raised. Specifically,  this problem setting consists of two independent studies sharing a common bivariate mean parameter $\bm\mu$, i.e., $\bm\mu_A=\bm\mu_B=\bm\mu$, where  $\bm\mu_A$ and $\bm\mu_B$ are the means of {\it Airbus} Study and {\it Boeing} Study, respectively. We construct a {\it depth-CD} from each study to carry out separately the two tests  $\bm\mu_A=\bm\mu_0$ and  $\bm\mu_B=\bm\mu_0$ with $\bm\mu_0=(15.85,432)^\prime$, and then combine the two test results using (\ref{eq:combining-special}) to draw the overall inference on testing the hypothesis in (\ref{hyp:FAA}).   

Specifically, we obtain a sample mean $\hat{\bm\mu}^*_A$ based on a bootstrap-$t$ sample of {\it Airbus} Study, and repeat this 2000 times to obtain a {\it depth-CD} $H_A(\bm\mu_A)$, in this case namely, the empirical distribution of $\{\hat{\bm\mu}^*_{A,1},\hat{\bm\mu}^*_{A,2},\ldots,\hat{\bm\mu}^*_{A,2000}\}$. A {\it depth-CD} $H_B(\bm\mu_B)$ for the {\it Boeing} study can be obtained similarly. We then combine $H_A(\bm\mu_A)$ and $H_B(\bm\mu_B)$ using (\ref{eq:combining-special}) for testing the hypothesis in (\ref{hyp:FAA}). Our fusion method yields a $p$-value of 0.008, indicating that the data provide strong evidence against the null hypothesis that the landings follow the FAA recommendation. This conclusion is drawn without assuming the sample follow a certain (say, normal) distribution. 

The seemingly contradictory results between the intuitive method and our fusion method may be best explained visually by the plots of individual {\it depth-CD}s for {\it Airbus} (blue circles) and {\it Boeing} (black crosses) in Figure~\ref{fig:FAA}. The {\it depth-CD}s here are represented by the empirical distributions of their respective bootstrap estimates. The red triangle marks the null value $\bm\mu_0=(15.85,432)$, which is clearly far from the centers of the two {\it depth-CD}s (which are the point estimates of their two respective means). The centrality values at $\bm\mu_0$ w.r.t. the two {\it depth-CD}s, or equivalently the two individual $p$-values, are 0.006 and 0.167. This finding implies low plausibilities for the assumption  $\bm\mu_A=\bm\mu_0$ or $\bm\mu_B=\bm\mu_0$. 
Thus, a small $p$-value (0.008) from our fusion method leading to the rejection of $H_0$ should be expected. 

We also plot in Figure~\ref{fig:FAA}  the {\it depth-CD} (green diamonds) obtained from the pooled data of the two studies. The red triangle $\bm\mu_0$ sits almost at the center of this {\it depth-CD}, which suggests that $\bm\mu_0$ as a plausible target value, as also reflected in a large $p$-value 0.956. This example shows that ignoring the heterogeneity of data sources or blindly aggregating data may mask important signals and lead to invalid and misleading conclusions.

Finally, to demonstrate the flexibility of our fusion method in handling more challenging situations, we suppose that the  recordings of the variable 'height' from {\it Airbus} aircraft are not available. In this scenario, traditional methods can make inference about 'height' only based on the landings of {\it Boeing} aircraft. For example, applying Hotelling's $T^2$ test to {\it Boeing} observations yields  a $p$-value of 0.152. This again yields an incorrect conclusion. Unlike traditional methods, our fusion method can efficiently incorporate  the information in the incomplete observations from {\it Airbus} Study, as shown in Section~\ref{sec:heter}. Combining the indirect evidence from {\it Airbus} with the direct evidence from {\it Boeing}, our method yields a $p$-value of 0.016. This result suggests strong evidence against the null hypothesis, which is consistent with our conclusion drawn from the complete data from both studies. Our analysis here shows that indirect evidence may contain valuable information (e.g., possibly through the correlation between 'height' and 'distance' in this case) without which incorrect inference outcome may be reached.

\section{Discussion}
We have used the concept of {\it depth-CD} and {\it depth-CV} to develop a new framework for fusion learning. This  fusion learning framework imposes no assumptions on the distribution of the data or statistics in each study. It has been shown to be efficient, general and robust by both theoretical properties and numerical studies. In the non-normal settings, it can reduce bias and improve efficiency in inference, as observed from simulation studies. In addition, our fusion framework can easily adapt to complex heterogeneous studies settings where existing methods fail. In particular, it can incorporate indirect evidence from heterogeneous studies for which the target parameter is not estimable, and achieve an additional gain of efficiency, as illustrated in both our simulation and case study. The phenomenon of incorporating indirect evidence to gain efficiency has also been observed, though in the normal or asymptotic normal settings, in e.g.,  \citep{yang2014efficient,liu2015multivariate,hoff2019smaller,chen2013using,chatterjee2016constrained,gao2017data}.
The last three combined  information from diverse studies through estimating equations, using large sample central limit theorem under parametric models.

The concept of {\it depth-CD} plays a key role in the development of our fusion framework. As a distribution function over the parameter space, it depicts ``confidence'' on each possible parameter value, in view of the given data. Intrinsically,  a {\it depth-CD} serves as a vehicle carrying all commonly used inference outcomes including point estimates, confidence intervals/regions, and $p$-values. This  all-encompassing characteristic affords our nonparametric combining scheme the desirable theoretical properties and good numerical performance seen in this paper. The concept of {\it depth-CD} is a natural extension of  {\it CD} to a general multivariate setting.  In the scalar or normal setting, the general {\it CD} has been proven a useful tool for solving other challenging problems in fusion learning. For instance, the idea of combining {\it CD}s leads to: robust inference with outlying studies \citep{xie2011confidence}, exact inference for discrete data \citep{liu2014exact,yang2016meta}, efficient inference for heterogeneous studies or network meta-analysis \citep{clagget2014meta,yang2014efficient,liu2015multivariate}, a split-conquer-combine approach for massive data \citep{chen2014split}, and individualized inference for a particular subject or study \citep{Shen2019}. See \cite{Cheng2017a} for a brief review on fusion learning via {\it CD}s.

A {\it depth-CV}, through {\it depth-CD}, is obtained by incorporating the idea of centrality measure from of data depth to construct nested central regions expanding with growing probability mass, in the setting of confidence distributions. The capturing of the nested central regions with their associate probability coverages is key in making {\it depth-CV} such a versatile and effective multivariate inference tool. This formulation of central regions expanding with growing probability is akin to those referred to as  ``quality index" and ``multivariate spacings" considered in  (\cite{liu1993quality}) and  (\cite{li2008multivariate}) in the context of assessing the underlying distribution for quality control purpose. The complete development of {\it depth-CV} and the fusion learning method in this paper may help broaden those two problem settings to make them more practical in reality, especially in multivariate control. 

Our fusion approach can facilitate the fusion of multivariate inferences from a wide range of data sources, which can be irregular, incomplete, or heterogeneous of various types. The development here may shed light on the possible extensions of {\it depth-CV} to the domains of directional data (data on circles/spheres) (\cite{liu1992ordering}) and functional data (\cite{claeskens2014multivariate,fan2019antipodal,lopez2009concept,  narisetty2016extremal}), where applications abound, including fusing existing different climate or weather forecast approaches. Those extensions may be worth exploring. Another useful extension could be in the direction of fusing related studies, as seen in in \cite{li2020sequential}.

\appendix
\section*{APPENDIX: PROOFS}
\begin{proof}[Proof of Lemma~\ref{lemma:centrality}]
  To prove Part (a), let $F(\cdot)$ denote the cumulative distribution function of $D_H(\bm\eta)$. It is a continuous function under the condition that $\pr_H\{\bm\eta: D_H(\bm\eta)=t\}=0$ for all $t$. Using the probability integral transform, we have $\pr_H\{\bm\eta:C_H(\bm\eta|D)\leq t\}=\pr_H\{\bm\eta:\pr_H\{\bm\xi: D_H(\bm\xi)\leq D_H(\bm\eta)\}\leq t\}=\pr_H\{\bm\eta: F(D_H(\bm\eta))\leq t\}=\pr_H\{\bm\eta:D_H(\bm\eta)\leq F^{-1}(t)\}=F(F^{-1}(t))=t$. Thus, $C_H(\bm\eta|D)$ follows the uniform distribution on [0,1]. For Part (b), the affine-invariance of the depth function directly leads to the affine-invariance of the corresponding centrality function.
\end{proof}

\begin{proof}[Proof of Proposition~\ref{prop:equiv}]
  It suffices to notice that Requirement (ii) that $\pr\{\bm\theta^o \in \mathcal{R}_{1-\alpha}(H_n(\bm X_n,\cdot))\}=1-\alpha, \forall \alpha \in (0,1)$, is equivalent to $\pr\{C_{H_n}(\bm\theta^o)\geq \alpha\}=1-\alpha$.
\end{proof}

\begin{proof}[Proof of Proposition~\ref{prop:point-est}]
  For any $\delta>0$,
  \begin{eqnarray*}
    \pr_{\bm\theta^o}\left\{||\hat{\bm\theta}_n^{MCE}-\bm\theta^o|| > \delta \right\} & = & \pr_{\bm\theta^o}\left\{ \left\{ ||\hat{\bm\theta}_n^{MCE}-\bm\theta^o|| > \delta \right\}\ \cap \left\{ \Delta_n(\epsilon) > \delta \right\} \right\}  \\
    & & +\ \pr_{\bm\theta^o}\left\{ \left\{ ||\hat{\bm\theta}_n^{MCE}-\bm\theta^o|| > \delta \right\}\ \cap \left\{ \Delta_n(\epsilon) \leq \delta \right\} \right\}
  \end{eqnarray*}
  The first term on the right hand side goes to zero as $n\to \infty$. The second term is no greater than $\pr_{\bm\theta^o}\{ C_{H_n(\bm X_n,\cdot)}(\bm\theta^o) \leq \epsilon \} = \epsilon$. Because $\epsilon$ is arbitrary, the left hand side of the above equation goes to zero as $n\to \infty$. Using similar arguments, we can prove that $\hat{\bm\theta}_n^{MCE} - \theta^o = O_p(a_n)$.
\end{proof}

\begin{proof}[Proof of Theorem~\ref{thm:CD-bootstrap}]
  The result of Theorem 3.1 (p.269) in \cite{liu1997notions} implies that $C_{B_n^*}(\bm\theta^o)$ converges in distribution to U(0,1) as $n \to \infty$. By Proposition~\ref{prop:equiv}, this proves that $B_n^*$ is a {\it depth-CD} for $\bm\theta$ asymptotically.
\end{proof}

\begin{proof}[Proof of Theorem~\ref{thm:CD-pivotal}]
  By Proposition~\ref{prop:equiv}, it suffices to show that $C_{\hat{\bm\theta}_n-A_n(\bm X_n)^{-1}\bm\eta_n}(\bm\theta^o)$ follows the uniform distribution on $(0,1)$. Making use of the result in Lemma~\ref{lemma:centrality}, we have
  \[
  \pr\{C_{\hat{\bm\theta}_n-A_n(\bm X_n)^{-1}\bm\eta_n}(\bm\theta^o) \leq t\} = \pr\{C_{\bm\eta_n}(A_n(\bm X_n)(\hat{\bm\theta}_n-\bm\theta^o)) \leq t\} = t.
  \]
  This completes the proof.
\end{proof}

\begin{lemma}
  Given {\it depth-CD}s $H_{k,n_k}(\bm\theta)$, $k=1,\ldots,K$, the statistic $C_{(c)}(\bm\theta^o)$, as a function of $K$ independent samples $\{X_{k,1},X_{k,2},\ldots,X_{k,n_k}\}$, follows the uniform distribution on [0,1], where $\bm\theta^o$ is the true value of $\bm\theta$.
  \label{lemma:combining}
\end{lemma}

\begin{proof}[Proof of Lemma~\ref{lemma:combining}]
  This is a direct consequence of Proposition~\ref{prop:equiv} and the definition of $G_c(\cdot)$.
\end{proof}

\begin{proof}[Proof of Theorem~\ref{thm:fused-CD}]
  Part(a) is a direct result of Lemma~\ref{lemma:combining}. Part(b) is due to the duality of hypothesis testing and confidence regions. We prove Part(c) by assuming that $\hat{\bm\theta}_{(c)}$ is not consistent. Without loss of generality, we set $n=n_1=\cdots=n_K$. Then, there exist constants $\delta_0 >0$ and $\rho_0>0$ such that for any integer $N_0>0$, there exist an integer $N>N_0$,
  \begin{equation}
  \label{eq:proof-thm-3-1}
  \pr_N \left\{ \| \hat{\bm\theta}_{(c)}-\bm\theta^o\| > \delta_0 \right\} > \rho_0.
  \end{equation}
  Let $\hat{\bm\theta}_k$ be the estimate obtained by maximizing the individual centrality function $C_{H_{k}}(\bm\theta)$. By Proposition~\ref{prop:point-est}, $\hat{\bm\theta}_k$ is consistent under the condition that $\Delta_{k,n}(\epsilon) \to 0$ for any $\epsilon>0$. Thus, when $N_0$ is sufficiently large, $
  \pr_N \left\{ \| \hat{\bm\theta}_k -\bm\theta^o \| > \delta_0/3 \ \right\} < \rho_0/(2K)
  $ for any $k$. This leads to the inequality below.
  \begin{equation}
  \label{eq:proof-thm-3-2}
  \pr_N \left\{ \| \hat{\bm\theta}_k -\bm\theta^o \| \leq \delta_0/3, k=1,2,\ldots,K \right\} > 1-\frac{\rho_0}{2}
  \end{equation}
  Combining the two inequalities in (\ref{eq:proof-thm-3-1}) and (\ref{eq:proof-thm-3-2}), we can show
  \begin{equation}
     \label{eq:proof-thm-3-3}
    \pr_N \left( A \right) = \pr_N \left( \left\{ \| \hat{\bm\theta}_k -\bm\theta^o \| \leq \delta_0/3, k=1,2,\ldots,K \right\} \bigcap  \left\{ \| \hat{\bm\theta}_{(c)}-\bm\theta^o\| > \delta_0 \right\}\right) > \frac{\rho_0}{2}.
  \end{equation}
  Other than a diminishing probability, the event $A$ implies that $C_{(c)}(\hat{\bm\theta}_1)>C_{(c)}(\hat{\bm\theta}_{(c)})$. This is concluded from observing the following three facts:

  (i) By definition, $C_{H_1}(\hat{\bm\theta}_1) \geq C_{H_1}(\hat{\bm\theta}_{(c)})$. Because $\|\hat{\bm\theta}_1 - \hat{\bm\theta}_{(c)} \| > 2\delta_0/3$ but  $\Delta_{1,n}(1) \to 0$ in probability, we conclude that $\pr_N \left\{ C_{H_1}(\hat{\bm\theta}_1) = C_{H_1}(\hat{\bm\theta}_{(c)}) \;\middle\vert\; A \right\} \to 0$, and thus,
  \begin{equation}
    \label{eq:proof-thm-3-4}
    \pr_N \left\{ C_{H_1}(\hat{\bm\theta}_1) > C_{H_1}(\hat{\bm\theta}_{(c)})  \;\middle\vert\; A \right\} \to 1.
  \end{equation}

  (ii) Now, we consider the case where $k \neq 1$. Due to the continuity of $C_{H_k}(\cdot)$, when $\delta_0$ is sufficiently small, we have $C_{H_k}(\bm\vartheta) \geq C_{H_k}(\hat{\bm\theta}_k) - \epsilon_0$ for any $\bm\vartheta \in S_{\delta_0/3}=\{ \bm\vartheta: \| \bm\vartheta - \bm\theta^o \| \leq \delta_0/3 \}$. Let $S^c_{\delta_0}=\{ \bm\vartheta: \| \bm\vartheta - \bm\theta^o \| > \delta_0 \}$. Since $\| \bm\vartheta-\bm\gamma \|>2\delta_0/3$ for any $\bm\vartheta \in S_{\delta_0/3}$ and $\bm\gamma \in S^c_{\delta_0}$, and $\Delta_{k,n}(\epsilon)\to 0$ in probability, we have $\pr_N \left\{ C_{H_k}(\bm\vartheta) > C_{H_k}(\bm\gamma),\ \text{for any}\ \bm\vartheta \in S_{\delta_0/3}\ \text{and}\  \bm\gamma \in S^c_{\delta_0}  \;\middle\vert\; A \right\} \to 1$. Notice that $\hat{\bm\theta}_1 \in S_{\delta_0/3}$ and $\hat{\bm\theta}_{(c)} \in S^c_{\delta_0}$, we obtain
  \begin{equation}
    \label{eq:proof-thm-3-5}
    \pr_N \left\{ C_{H_k}(\hat{\bm\theta}_1) > C_{H_k}(\hat{\bm\theta}_{(c)})  \;\middle\vert\; A \right\} \to 1.
  \end{equation}

  (iii) Given that the function $g_c(u_1,u_2,\ldots,u_K)$ is increasing in each of its coordinates, (\ref{eq:proof-thm-3-4}) and (\ref{eq:proof-thm-3-5}) lead to
  \begin{equation}
  \label{eq:proof-thm-3-6}
  \pr_N \left\{ C_{(c)}(\hat{\bm\theta}_1)>C_{(c)}(\hat{\bm\theta}_{(c)}) \;\middle\vert\; A \right\} \to 1.
  \end{equation}
  As a result of (\ref{eq:proof-thm-3-3}) and (\ref{eq:proof-thm-3-6}), we establish that
  \[
  \pr_N \left\{ C_{(c)}(\hat{\bm\theta}_1)>C_{(c)}(\hat{\bm\theta}_{(c)}) \right\} \geq \pr_N \left( A \right) -\frac{\rho_0}{4} > \frac{\rho_0}{4}.
  \]
  In other words, there is a non-zero probability that $\hat{\bm\theta}_{(c)}$ is not the maximizer of the function $C_{(c)}(\bm\theta)$, which contradicts the definition of $\hat{\bm\theta}_{(c)}$. This completes the proof.
\end{proof}

\begin{proof}[Proof of theorem~\ref{thm:accuracy}]
  For notational simplicity, we prove the result for the case $K=2$. Making use of the conditional expectation, we can establish
  \begin{eqnarray*}
    \pr\left\{C_{(c)}(\bm\theta^o) \leq a \right\} & = & \pr\left\{ G_c\left(g_c(C_{H_{1,n_1}}(\bm\theta^o),C_{H_{2,n_2}}(\bm\theta^o))\right) \leq a \right\}\\
    & = & E \left[ \pr\left\{ G_c\left(g_c(C_{H_{1,n_1}}(\bm\theta^o),C_{H_{2,n_2}}(\bm\theta^o))\right) \leq a  \right\}  \;\middle\vert\; C_{H_{2,n_2}}(\bm\theta^o) \right]\\
    & = & E \left[ \pr\left\{ C_{H_{1,n_1}}(\bm\theta^o) \leq g_c^{-1}\left[ G_c^{-1}(a)\;\middle\vert\; C_{H_{2,n_2}}(\bm\theta^o) \right] \right\}  \;\middle\vert\; C_{H_{2,n_2}}(\bm\theta^o) \right]\\
    & = & E \left( g_c^{-1}\left[ G_c^{-1}(a)\;\middle\vert\; C_{H_{2,n_2}}(\bm\theta^o) \right] \right) + O(n^{-j/2})\\
    & = & E \left[ \pr\left\{ U_1 \leq g_c^{-1}\left[ G_c^{-1}(a)\;\middle\vert\; C_{H_{2,n_2}}(\bm\theta^o) \right] \right\}  \;\middle\vert\; C_{H_{2,n_2}}(\bm\theta^o) \right] + O(n^{-j/2}).\\
  \end{eqnarray*}
  Here, $U_1$ (and $U_2$ below) is a random variable following the uniform distribution on [0,1] and independent of the other variables. We re-write the right-hand side of the above equation and obtain
  \begin{eqnarray*}
    \pr\left\{C_{(c)}(\bm\theta^o) \leq a \right\} & = & \pr\left\{ G_c\left(g_c(U_1,C_{H_{2,n_2}}(\bm\theta^o))\right) \leq a \right\} + O(n^{-j/2}).
  \end{eqnarray*}
  Conditioning on $U_1$ and using similar derivations, we have
  \begin{eqnarray*}
    \pr\left\{C_{(c)}(\bm\theta^o) \leq a \right\} & = & \pr\left\{ G_c\left(g_c(U_1,U_2)\right) \leq a \right\} + O(n^{-j/2}) + O(n^{-j/2})\\
    & = & a  + O(n^{-j/2}).
  \end{eqnarray*}
  This completes the proof.
\end{proof}

\begin{lemma}
  \label{lemma:bahadur}
  Let $C_{(c)}(u_1,u_2,\ldots,u_K)$ be a mapping from $[0,1]^K$ to $[0,1]$. Assume that it is increasing in each coordinate and that $C_{(c)}(U_1,U_2,\ldots,U_K)$ follows the $U[0,1]$ distribution when $U_1,U_2,\ldots,U_K$ are independent random variables following the $U[0,1]$ distribution. Then, the inequality $C_{(c)}(u_1,u_2,\ldots,u_K) \geq \prod_{k=1}^K u_k$ holds for any $u_1,u_2,\ldots,u_K \in [0,1]$.
\end{lemma}

\begin{proof}[Proof of Lemma~\ref{lemma:bahadur}]
  The proof can be found in \cite{singh2005combining} (p.180).
\end{proof}

\begin{proof}[Proof of Theorem~\ref{thm:bahadur}]
  To prove the inequality in (\ref{eq:bahadur-bound}), we take $u_k=C_{H_{k,n_k}}(\bm\theta^o+b\bm\lambda)$ in Lemma~\ref{lemma:bahadur}. Then, it follows that
  \[
  C_{(c)}(\bm\theta^o + b\bm\lambda) \geq \prod_{k=1}^K C_{H_{k,n_k}}(\bm\theta^o+b\bm\lambda).
  \]
  Rearrangement of the above inequality leads to
  \[
    - \log \{C_{(c)}(\bm\theta^o+b \bm\lambda) \} /n \leq - \sum_{k=1}^K \log \{C_{H_{k,n_k}}(\bm\theta^o+b \bm\lambda) \} /n
  \]
  Taking the limit superiors of both sides results in the desired inequality in (\ref{eq:bahadur-bound}).

  To prove the equality in (\ref{eq:bahadur-E}), we write
  \begin{eqnarray*}
     & & -\frac{1}{n}\log \{C_{(c)}(\bm\theta^o+b \bm\lambda) \}\\
     & = &  -\frac{1}{n}\log \left( \pr \left\{\chi_{2K}^2 \geq -2 \sum_{k=1}^K \log(C_{H_{k,n_k}}(\bm\theta^o+b\bm\lambda)) \right\} \right)\\
     & = & \frac{2}{n}\sum_{k=1}^K \log(C_{H_{k,n_k}}(\bm\theta^o+b\bm\lambda)) \cdot \frac{\log \left( \pr \left\{\chi_{2K}^2 \geq -2 \sum_{k=1}^K \log(C_{H_{k,n_k}}(\bm\theta^o+b\bm\lambda)) \right\} \right)}{-2 \sum_{k=1}^K \log(C_{H_{k,n_k}}(\bm\theta^o+b\bm\lambda))}.
  \end{eqnarray*}
  Making use of the fact that $\lim_{t \to \infty}\frac{1}{t} \log \pr \left\{ \chi^2_{2K} \geq t \right\} = -\frac{1}{2}$, the right side of the above equation converges to $-2\sum_{k=1}^K a_k S_{k,\bm\lambda}(b)\cdot (-\frac{1}{2})=\sum_{k=1}^K a_k S_{k,\bm\lambda}(b)$ when $n \to \infty$.
\end{proof}

\begin{proof}[Proof of Theorem~\ref{thm:fused-CD-heter}]
  Lemma~\ref{lemma:combining} still holds in the setting of Theorem~\ref{thm:fused-CD-heter}. Therefore, one can easily conclude the results in Parts(a)-(b). The result in Part(c) can be established in a similar way as we prove Part(c) of Theorem~\ref{thm:fused-CD}. The details are thus omitted.
\end{proof}

\bibliographystyle{asa}
\bibliography{bibli_nonparametric}

\newpage

\begin{sidewaysfigure}
\includegraphics[width=1\textwidth]{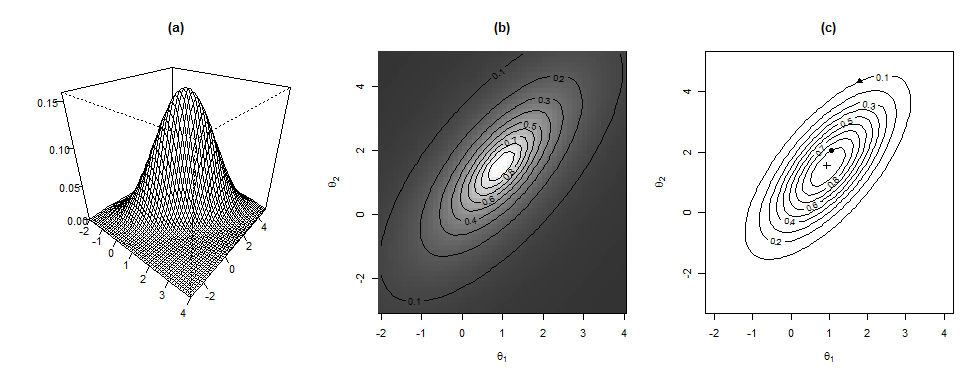}
    \caption{Illustrations of a {\it depth-CD} for the mean parameter $\bm\theta$ in $BN(\bm\theta, \Sigma)$: (a) a 3D-surface plot for the {\it depth-CD}; (b) a grey-color heat map for the depth contours with Mahalanobis depth values. The contours in (c) in the parameter space connect the parameter values of the same centrality value.  (c) illustrates the utility of the {\it depth-CV} for drawing confidence regions, $p$-values, and a point estimate. The plots are based on a simulated sample of size  $n=20$ and $\bm\theta=(1,1)$.} 
        \label{fig:CD-Bivariate-Normal}
\end{sidewaysfigure}

\begin{figure}
    \includegraphics[width=1\textwidth]{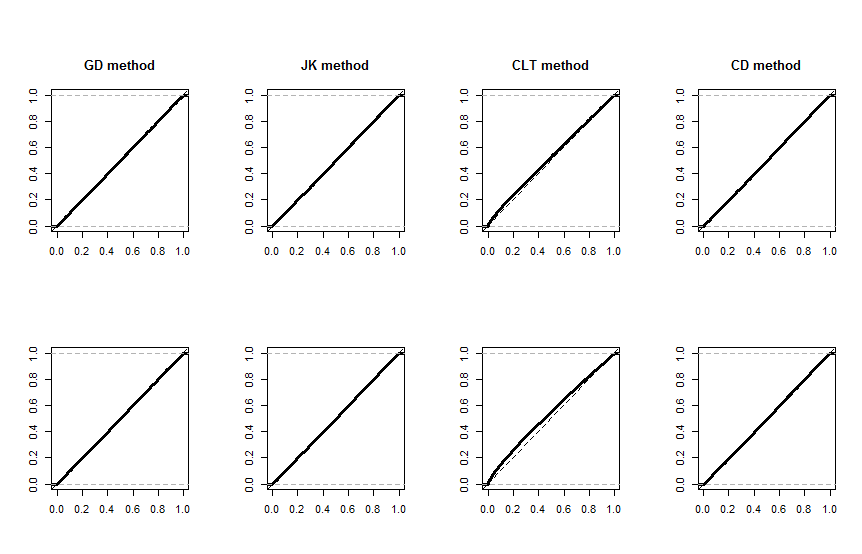}
    \caption{The null distributions of $p$-values derived from an individual study (upper row) and from the combined inference (lower row) for the common mean. The sample of size $n=30$ in each individual study are drawn from a bivariate normal distribution.}
    \label{fig:CD-combining-normal}
\end{figure}

\begin{figure}
    \includegraphics[width=1\textwidth]{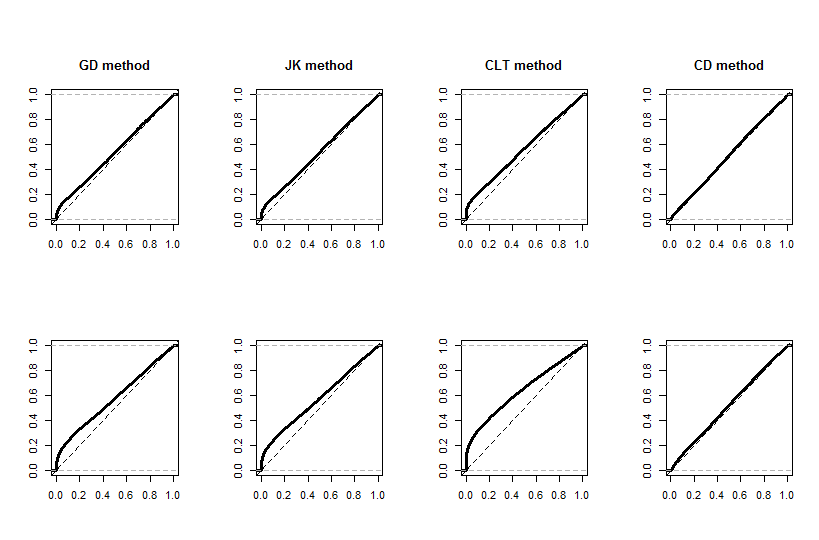}
    \caption{The null distributions of $p$-values derived from an individual study (upper row) and from the combined inference (lower row) for the common mean. The sample of size $n=30$ in each individual study are drawn from a bivariate $\chi^2$ distribution.}
    \label{fig:CD-combining-chisq}
\end{figure}

\begin{figure}
    \includegraphics[width=1\textwidth]{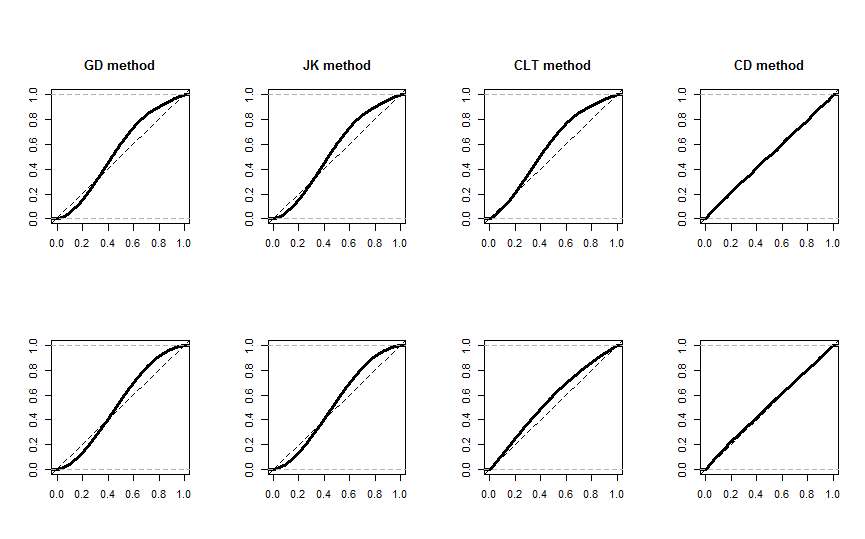}
    \caption{The null distributions of $p$-values derived from an individual study (the upper row) and the combined inference (the lower row) for the common mean. The sample of size $n=30$ in each individual study are drawn from a bivariate Cauchy distribution.}
    \label{fig:CD-combining-cauchy}
\end{figure}

\begin{figure}
    \includegraphics[width=1\textwidth]{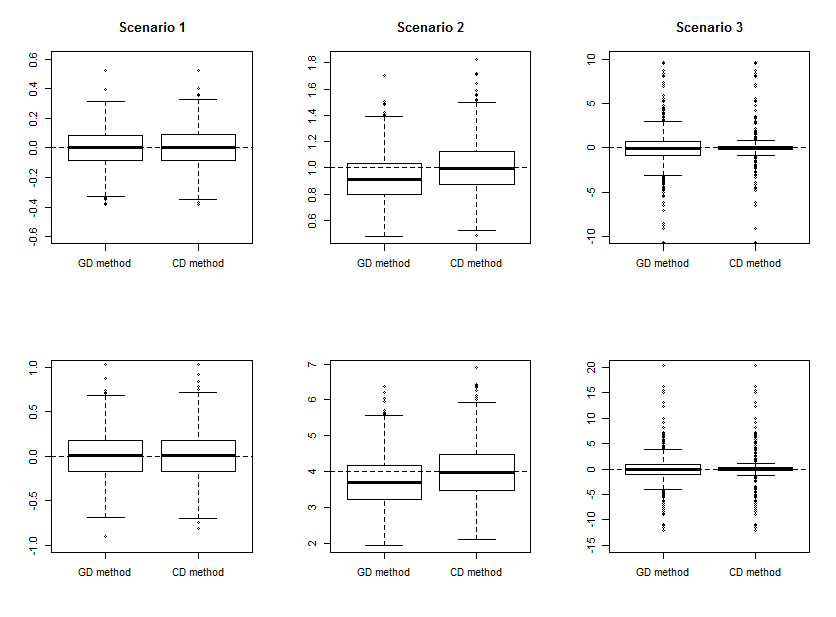}
    \caption{Boxplots of Greybill-Deal estimates and CD estimates for inferring the common mean (or location) vector $\bm\mu=(\mu_1,\mu_2)^\prime$ (the upper row for $\mu_1$ and the lower row for $\mu_2$). The true values $\bm\mu_0$ are drawn in the dashed lines. The sample of size $n=30$ in each individual study are drawn from Scenario 1 ( bivariate normal), Scenario 2 ($\chi^2$), and Scenario 3 (Cauchy).}
    \label{fig:estimate}
\end{figure}

\begin{figure}
    \includegraphics[width=1\textwidth]{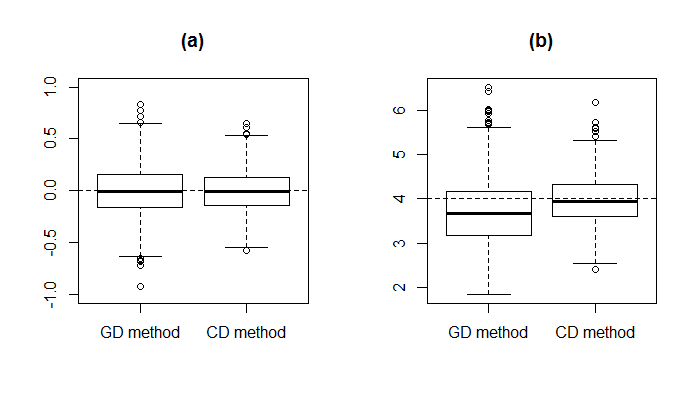}
    \caption{Boxplots of Greybill-Deal estimates and CD estimates for inferring $\mu_2$ in the common mean vector $\bm\mu=(\mu_1,\mu_2)^\prime$. The underlying distribution of $\bm X$ is bivariate normal in (a) and $\chi^2$ in (b), with the true value of $\mu_2$ being 0 and 4, respectively (drawn in the dashed lines). The sample size  in each individual study is $30$.}
    \label{fig:estimate-hetero}
\end{figure}

\begin{figure}
    \includegraphics[width=1\textwidth]{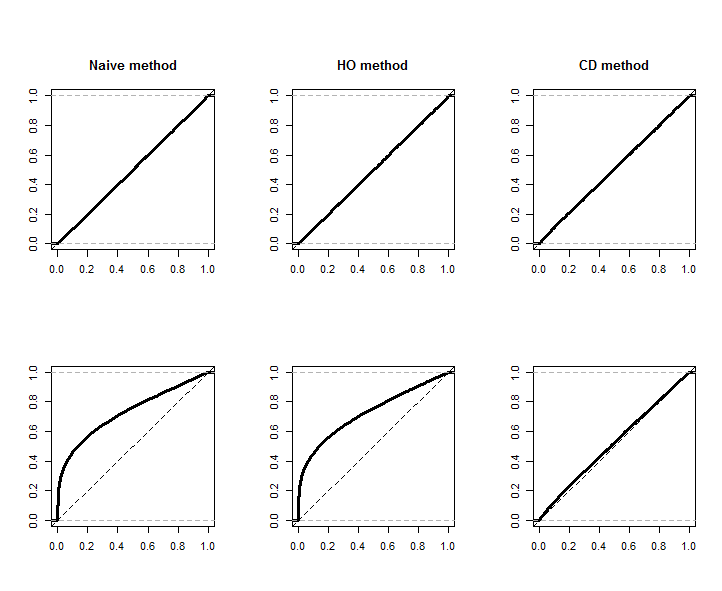}
    \caption{The null distributions of $p$-values for meta-analysis of correlation coefficients. The sample of size $n=200$  in each individual study are drawn from a bivariate normal distribution (upper row) or a bivariate normal-$\chi^2$ distribution (lower row).}
    \label{fig:corr}
\end{figure}

\begin{figure}
    \center\includegraphics[width=0.6\textwidth]{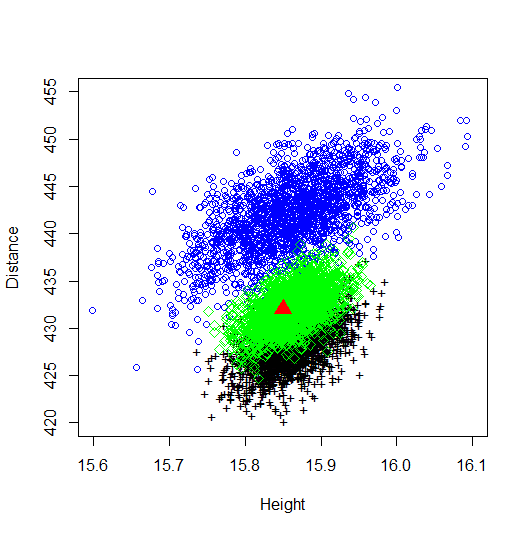}
    \caption{{\it depth-CD}s for each individual study, {\it Airbus} (blue circles) and {\it Boeing} (black crosses), and a depth-CD by aggregating data from the two studies as if they were from the same source (green diamonds). The {\it depth-CD}s here are represented by the empirical distributions of bootstrap estimates. The red triangle indicates the target value $\bm\mu_0$ in the null hypothesis $H_0: \bm\mu_A=\bm\mu_B=\bm\mu_0=(15.85,432)^\prime$.}
    \label{fig:FAA}
\end{figure}

\begin{table}[ht]
\small
\caption{{Empirical distribution of the $p$-values at the null for the common mean problem}}
\vspace{0.2in}
\begin{tabular*}{\textwidth}{@{\extracolsep{\fill}}l*{10}{r}}
  \toprule
  \multicolumn{11}{l}{Scenario 1. (Normal distribution)}\\
  \midrule
 Nominal Probs & 0.05 & 0.1 & 0.2 & 0.3 & 0.4 & 0.5 & 0.6 & 0.7 & 0.8 & 0.9 \\
  \midrule
GD method & 0.05 & 0.10 & 0.20 & 0.30 & 0.39 & 0.49 & 0.60 & 0.70 & 0.80 & 0.90 \\
  JK method & 0.05 & 0.10 & 0.20 & 0.30 & 0.39 & 0.49 & 0.60 & 0.70 & 0.80 & 0.90 \\
  CLT method & \bf 0.09 & \bf 0.15 & 0.26 & 0.36 & 0.46 & 0.55 & 0.65 & 0.74 & 0.83 & 0.91 \\
  CD method & 0.04 & 0.10 & 0.19 & 0.29 & 0.39 & 0.50 & 0.60 & 0.70 & 0.80 & 0.90 \\
  \midrule
   \multicolumn{11}{l}{Scenario 2. ($\chi^2$ distribution)}\\
   \midrule
   GD method & \bf 0.18 & \bf 0.24 & \bf 0.33 & \bf 0.41 & \bf 0.49 & 0.58 & 0.66 & 0.75 & 0.84 & 0.92 \\
  JK method & \bf 0.18 & \bf 0.24 & \bf 0.33 & \bf 0.41 & \bf 0.49 & 0.58 & 0.66 & 0.75 & 0.84 & 0.92 \\
  CLT method & \bf 0.25 & \bf 0.32 & \bf 0.42 & \bf 0.50 & \bf 0.59 & \bf 0.66 & \bf 0.73 & \bf 0.80 & 0.87 & 0.94 \\
  CD method & 0.07 & 0.12 & 0.23 & 0.32 & 0.42 & 0.52 & 0.62 & 0.72 & 0.82 & 0.92 \\
  \midrule
  \multicolumn{11}{l}{Scenario 3. (Cauchy distribution)}\\
   \midrule
  GD method & \bf 0.01 & 0.04 & 0.14 & 0.26 & 0.40 & 0.56 & \bf 0.70 & \bf 0.82 & \bf 0.91 & 0.97 \\
  JK method & \bf 0.01 & 0.04 & 0.14 & 0.26 & 0.40 & 0.56 & \bf 0.70 & \bf 0.82 & \bf 0.91 & 0.97 \\
  CLT method & 0.05 & 0.12 & 0.25 & 0.37 & \bf 0.49 & \bf 0.60 & \bf 0.69 & 0.78 & 0.86 & 0.93 \\
  CD method & 0.06 & 0.12 & 0.22 & 0.32 & 0.42 & 0.52 & 0.61 & 0.71 & 0.80 & 0.90 \\
  \bottomrule
\end{tabular*}
{\it Remark:} Boldfaced are those values with notable deviations from the nominal value.
\label{tab:ecdf-pvalue}
\end{table}

\end{document}